\def\BE{\begin{equation}}
\def\EE#1{\label{#1}\end{equation}}
\def\be{\begin{align}}
\def\se#1{\begin{subequations}\label{#1}
\renewcommand{\theequation}{\theparentequation.\arabic{equation}}}
\def\UP{\rule[1em]{0ex}{.5ex}}
\def\DO{\rule[-.5ex]{0ex}{1ex}}
\def\R{{\mathbb R}}
\def\T{{\mathbb T}}
\def\rf#1{(\ref{#1})}
\def\xf#1{Fig.~\ref{#1}}
\def\i{{\rm i}}
\def\e{{\rm e}^{\i\bf q\cdot X}}
\def\d{{\rm d}}
\def\L{\boldsymbol{\mathfrak{L}}}
\def\A{\mathfrak{A}}
\def\At{\boldsymbol{\mathfrak{A}}}
\def\Z{\boldsymbol{\mathfrak{Z}}}
\def\z{\mathfrak{z}}
\def\D{\mathfrak{D}}
\def\Db{\boldsymbol{\mathfrak{D}}}
\def\Dt{\widetilde\D}
\def\Dw{\widetilde\Db}
\def\B{\mathfrak{B}}
\def\ss{{\bf s}^{\rm s}}
\def\sa{{\bf s}^{\rm a}}
\def\Zs{{\bf Z}^{\rm s}}
\def\Za{{\bf Z}^{\rm a}}
\def\rZ{{\rm^r}{\bf Z}}
\def\As{\At^{\rm s}}
\def\Aa{\At^{\rm a}}
\def\LA{\left\langle}
\def\RA{\right\rangle}
\def\fs{\varphi_{\rm s}}
\documentclass[12pt,a4paper]{Melsarticle}
\usepackage{amsmath,amssymb,graphicx,bm,subcaption,multirow}
\binoppenalty=\maxdimen
\relpenalty=\maxdimen
\begin{document}
\title{Negative magnetic eddy diffusivity due to oscillogenic $\alpha$-effect}
\author[mitp]{A.~Andrievsky}
\author[por]{R.~Chertovskih}
\author[mitp]{V.~Zheligovsky}
\address[mitp]{Institute of Earthquake Prediction Theory and
Mathematical Geophysics, Russian Ac. Sci.,\\
84/32 Profsoyuznaya St, 117997 Moscow, Russian Federation}
\address[por]{Research Center for Systems and Technologies, Faculty of
Engineering, University of Porto,\\
Rua Dr.~Roberto Frias, s/n, 4200-465, Porto, Portugal}

\begin{abstract}
We study large-scale kinematic dynamo action of steady mirror-antisymmetric
flows of incompressible fluid, that involve small spatial scales only,
by asymptotic methods of the multiscale stability theory. It turns out that,
due to the magnetic $\alpha$-effect in such flows, the large-scale mean field
experiences harmonic oscillations in time on the scale O($\varepsilon t$)
without growth or decay. Here $\varepsilon$ is the spatial scale ratio and $t$
is the fast time of the order of the flow turnover time. The interaction
of the accompanying fluctuating magnetic field with the flow gives rise
to an anisotropic magnetic eddy diffusivity, whose dependence
on the direction of the large-scale wave vector generically exhibits a singular
behaviour, and thus to negative eddy diffusivity for whichever molecular
magnetic diffusivity. Consequently, such flows always act as kinematic dynamos
on the time scale O($\varepsilon^2t$); for the directions at which eddy
diffusivity is infinite, the large-scale mean-field growth rate is finite
on the scale O($\varepsilon^{3/2}t$). We investigate numerically this
dynamo mechanism for two sample flows.
\end{abstract}
\maketitle

\section{Introduction}\label{intr}

The multiscale stability theory (MST) examines generation of large-scale
magnetic field by a small-scale flow in the limit of high scale separation, for
the spatial scale ratio (henceforth denoted by~$\varepsilon$) presumed to be
infinitesimally small. While the scope of MST is narrower than that
of the mean-field electrodynamics (see a detailed discussion in \cite{ABNZ} and
references therein), all MST results are obtained
by asymptotic methods from the first principles (the magnetic induction
equation, when kinematic dynamo is under scrutiny) without recourse
to additional assumptions (such as the validity of the second-order correlation
approximation, SOCA, sometimes used in the mean-field electrodynamics).

MST establishes (see \cite{VZ} and references therein) that in a two-scale
space-periodic kinematic dynamo the magnetic $\alpha$-effect and eddy
diffusivity never act simultaneously as predominant mechanisms
for large-scale field generation. They run on different time scales: either
an $\alpha$-effect dynamo operates on the so-called {\it slow} time
$T_1=\varepsilon t$, or the negative magnetic eddy diffusivity does this
on the slow time $T_2=\varepsilon^2 t$ (or, of course,
there can be no large-scale generation at all). Here $t$ is the {\it fast} time
of the order of the small-scale flow turnover time; fields depending solely
on the {\it fast} spatial variable $\bf x$ are called {\it small-scale},
while {\it large-scale} fields also depend on the {\it slow} variable
$\bf X=\varepsilon x$. The scale ratio $\varepsilon$ is a small parameter,
which gives an opportunity to use asymptotic techniques for homogenisation
of elliptic operators. Magnetic modes (i.e.,
eigenfunctions of the magnetic induction operator) considered here have
the structure of space-periodic small-scale fields that are amplitude-modulated
by the large-scale Fourier harmonics $\e$.

In the presence of some of the two effects, the growth rates of large-scale
magnetic modes are controlled (up to higher-order, in $\varepsilon$, terms)
by the spectrum of the $\alpha$-effect or eddy diffusivity operators,
respectively. The spectrum of the $\alpha$-effect operator
$\LA{\bf h}\RA\mapsto\nabla\times\At\LA\bf h\RA$, acting on space-periodic mean
fields $\LA\bf h\RA$, is symmetric about the imaginary axis \cite{Vi} (here
$\At$ is the $\alpha$-effect tensor, $\At\LA\bf h\RA$ being the mean
electromotive force arising due to the interaction of the small-scale flow and
the small-scale components of the magnetic field; angle brackets denote
averaging, we will define the appropriate averaging in the next section).
For a given wave vector $\bf q$, the eigenvalues of the $\alpha$-effect and
eddy diffusivity operators are proportional to $|\bf q|$ and $|{\bf q}|^2$,
respectively. A generic mean magnetic
field involves infinitely many large-scale magnetic modes for wave vectors
$\bf q$ of increasing length; it thus grows superexponentially in the slow times $T_1$ or
$T_2$ under the action of the $\alpha$-effect and negative eddy diffusivity,
respectively. The growing magnetic field increasingly perturbs the flow via
the Lorentz force; this affects the field generation. Consequently,
the magnetic $\alpha$-effect is a relatively rapidly self-destructing mechanism
for large-scale generation (see \cite{ChZh}), and it can be
of (astro)physical significance only while remaining weak -- ideally,
just causing temporal oscillations of the mean magnetic field, which happens
when all eigenvalues of the $\alpha$-effect operator are imaginary. We call
{\it oscillogenic} an $\alpha$-effect that yields constant-amplitude
harmonic oscillations in time of the mean magnetic field which has initially
the spatial profile of a Fourier harmonics. We note that this is a property
of the $\alpha$-effect and not of the flow, because the flows considered here,
that give rise to such an $\alpha$-effect, are steady and at least some of them
can kinematically generate small-scale growing magnetic fields for sufficiently
small molecular diffusivities. Applying MST tools, we will examine here
the joint action of an oscillogenic $\alpha$-effect and the magnetic eddy
diffusivity.

The following symmetry is relevant for our constructions. A vector field
${\bf f}=(f_1,f_2,f_3)$
is called {\it symmetric in a Cartesian variable} $x_i$, if
$$f_j((-1)^{\delta^i_1}x_1,(-1)^{\delta^i_2}x_2,(-1)^{\delta^i_3}x_3)
=(-1)^{\delta^i_j}\,f_j({\bf x})$$
for all $i$ and $j$ (such that $1\le i,j\le3$), and {\it antisymmetric in}
$x_i$, if
\BE f_j((-1)^{\delta^i_1}x_1,(-1)^{\delta^i_2}x_2,(-1)^{\delta^i_3}x_3)
=(-1)^{1-\delta^i_j}\,f_j({\bf x})\EE{as}
for all $i$ and $j$. Here $\delta^i_j$ is the Kronecker symbol.
A field $\bf f$ is called {\it parity-invariant}, if
$$\bf f(x)=-f(-x).$$
When a flow is symmetric in all $x_i$, then it is parity-invariant.
Parity-invariant flows do not give rise to the $\alpha$-effect, the dominant
large-scale effect that they can sustain is the magnetic eddy diffusivity.
The symmetry and antisymmetry in a Cartesian variable, as well as parity
invariance are compatible with the solenoidality of a vector field.

The combination of the oscillogenic $\alpha$-effect and eddy diffusivity
was not considered in \cite{VZ} on the grounds that flows giving rise to
the oscillogenic $\alpha$-effect are non-generic. However, as we show in section
\ref{osc}, the oscillogenic $\alpha$-effect is encountered in any steady flow
antisymmetric in a Cartesian coordinate. The antisymmetry and parity invariance
are both defined by how field components are transformed under
the reversal of some Cartesian coordinate axes; in both cases, the number
of such relations is equal to the dimension of the space. Consequently, flows
featuring the oscillogenic magnetic $\alpha$-effect are not ``less generic''
than flows in which magnetic eddy diffusivity is the dominant large-scale
effect, and therefore dynamos powered by flows possessing
such an $\alpha$-effect equally deserve to be investigated.

We may note that, unlike parity invariance and the symmetry in a Cartesian
coordinate, the antisymmetry in a coordinate is incompatible with the dynamical
equations of fluid motion (the Euler or Navier--Stokes equations), i.e.,
a steady flow symmetric in $x_i$ persists only under a suitable forcing.
However, MST is meant to explore the interaction of just two significantly
different scales within the entire hierarchy of spatial scales; the necessary
forcing can then be supplied by the interaction of scales that are outside
the scope of the multiscale formalism.

The paper is organised as follows. We briefly recall how the magnetic
$\alpha$-effect operator is derived by the multiscale techniques in section
\ref{mul} and study its spectrum for steady flows possessing the antisymmetry under
consideration in section \ref{osc}. In section \ref{hi} we discuss symmetry
properties of the magnetic eddy diffusivity tensor for such flows, and
establish that, due to interaction with the $\alpha$-effect, eddy diffusivity
is guaranteed to be negative and has a singularity. In section \ref{nr} we
perform a numerical investigation of the large-scale dynamo for two flows
possessing the required antisymmetry. Finally, in section \ref{sng} we
consider wave vectors for which eddy diffusivity is singular, derive
alternative asymptotic expansions for large-scale magnetic modes and their
growth rates, and show that in this case the dynamo is faster, growth rates
being finite in the slow time $T_{3/2}=\varepsilon^{3/2}t$ instead of $T_2$
otherwise.

We consider hydromagnetic dynamo for steady flows of incompressible fluid,
whose magnetic permeability is constant, in the absence of any additional
sources of the field. Magnetic field generation is studied
in the kinematic regime, i.e., we consider a linear problem for an elliptic
operator; the problem is homogeneous in magnetic field. We undimensionalise
all physical fields and quantities. In computations, flows are normalised,
so that the r.m.s.~flow velocity is unity. Consequently, our molecular
diffusivity can serve as the inverse magnetic Reynolds number.

\section{The multiscale formalism}\label{mul}

For reader's convenience, we now briefly outline the standard
multiscale formalism describing two-scale kinematic dynamos (see, e.g.,
\cite{VZ}). The assumed antisymmetry of the flow affects
the structure of the $\alpha$-effect tensor $\At$ (see section~\ref{osc})
as well as comes into play when we consider the third set of equations
(see section~\ref{hi}) in the hierarchy derived in MST.

\subsection{The approach}
The evolution of a magnetic field $\bf h$ in a volume of a conducting fluid
is governed by the equation
\BE{\partial{\bf h}\over\partial t}=\L{\bf h},\EE{mie}
where
\BE\L{\bf h}=\eta\nabla^2{\bf h}+\nabla\times({\bf v}\times{\bf h})\EE{Lde}
is the magnetic induction operator, $\bf v$ the flow velocity and $\eta$
the magnetic molecular diffusivity. For the sake of simplicity, we will
consider steady flows $\bf v(x)$ (although it must be noted that the algebra
remains virtually unchanged for flows, periodic in time). The kinematic dynamo
problem can then be formulated as the eigenvalue problem
\BE\L{\bf h}=\lambda{\bf h}.\EE{Main}

The magnetic mode $\bf h(x,X)$ is supposed to depend on both the fast and slow
spatial variables. We assume that $\bf v(x)$
and $\bf h(x,X)$ are $2\pi$-periodic in the fast variables $x_i$; the two
fields are solenoidal; $\bf v(x)$ is small-scale and zero-mean,
$\LA{\bf v}\RA=0$. The spatial mean over the periodicity cell $\T^3$
in the fast spatial variables and the fluctuating part of a field are defined
by the relations
$$\LA{\bf f(x,X)}\RA={1\over(2\pi)^3}\int_{\T^3}{\bf f}({\bf x,X})\,\d{\bf x}
=\sum_{k=1}^3\LA{\bf f}\RA_k{\bf e}_k,\qquad\{{\bf f}\}={\bf f}-\LA{\bf f}\RA,$$
where ${\bf e}_k$ are unit vectors of the Cartesian coordinate system.
Differential operators acting on the slow spatial variables will be decorated
with the subscript $\bf X$, and non-decorated ones will denote the respective
differential operations in the fast variables; the magnetic induction operator
\rf{Lde}, $\L$, is henceforth supposed to act on the fast variables only.

The kernel of the operator, adjoint to the magnetic induction operator,
\BE\L^*{\bf h}=\eta\nabla^2{\bf h}-{\bf v}\times(\nabla\times{\bf h}),\EE{La}
involves constant vector fields and hence its dimension is at least three;
generically, $\dim\ker\L^*=3$. We assume that the pair ($\eta,\bf\ v(x)$)
is generic, i.e., the kernel of $\L^*$ consists of constant fields.
Using the Fredholm alternative theorem \cite{Lus}, we then can show
(see \cite{VZ}) that the condition $\LA{\bf f}\RA=0$ is necessary and
sufficient for the solvability of the equation $\L\bf h=f$ (spatial averaging
of the equation delivers a straightforward demonstration that the condition
is necessary).

We seek solutions to \rf{Main} as power series in $\varepsilon$:
\se{ps}\be{\bf h(x,X})=&\sum_{n=0}^{\infty}{\bf h}_n({\bf x,X})\varepsilon^n,\label{h}\\
\lambda=&\sum_{n=0}^{\infty}\lambda_n\varepsilon^n.\label{e}\end{align}\end{subequations}
Substituting \rf{ps} into \rf{Main} we obtain
\BE\sum_{n=0}^\infty\left(\L{\bf h}_n
+\eta(2(\nabla\cdot\nabla_{\bf X}){\bf h}_{n-1}+\nabla_{\bf X}^2{\bf h}_{n-2})
+\nabla_{\bf X}\times({\bf v}\times{\bf h}_{n-1})
-\sum_{m=0}^n\lambda_{n-m}{\bf h}_m\right)\varepsilon^n=0\EE{hie}
(by definition, ${\bf h}_n=0$ for $n<0$).
The solenoidality of the magnetic mode implies relations
\be\nabla_{\bf X}\cdot\LA{\bf h}_n\RA&=0,\label{so}\\
\nabla\cdot{\bf h}_n+\nabla_{\bf X}\cdot\{{\bf h}_{n-1}\}&=0.\nonumber\end{align}
that hold for all $n\ge0$.

\subsection{Order $\varepsilon^0$ equation}\label{O1}
For $n=0$, we deduce from \rf{hie} the equation
\BE\L{\bf h}_0=\lambda_0{\bf h}_0.\EE{eq_0}
Averaging it yields $0=\lambda_0\LA{\bf h}_0\RA$. Since our goal
is to explore large-scale dynamos, we select the possibility $\lambda_0=0$
(another potentially interesting case occurring for an imaginary
$\lambda_0\ne0$ is not considered here since it is not generic).
For Re\,$\lambda_0\ne 0$, \rf{h} is just a large-scale perturbation
of the small-scale mode associated with the eigenvalue $\lambda_0$, and
the underlying mechanism for generation is small-scale.

By linearity of $\L$, we now find from \rf{eq_0}
\BE{\bf h}_0=\sum_{k=1}^3\LA{\bf h}_0\RA_k{\bf s}_k,\EE{so0}
where neutral magnetic modes ${\bf s}_k({\bf x})$
are solutions to {\it auxiliary problems of type I}:
\BE\L{\bf s}_k=0,\qquad\LA{\bf s}_k\RA={\bf e}_k,\qquad
\nabla\cdot{\bf s}_k=0.\EE{Seq}
Existence of the modes follows from that the kernels of $\L^*$ and $\L$ have
the same dimension, and eigenfunctions of an elliptic operator ($\L$ in our
case) comprise a basis in the Lebesgue space L$_2(\T^3)$ (see~\cite{Ar,VZ}).

\subsection{Order $\varepsilon^1$ equation}
For $n=1$, \rf{hie} implies
\BE\L{\bf h}_1+2\eta(\nabla\cdot\nabla_{\bf X}){\bf h}_0
+\nabla_{\bf X}\times({\bf v}\times{\bf h}_0)=\lambda_1{\bf h}_0.\EE{eq_1}
Substituting \rf{so0} and averaging this equation, we find
the condition for its solvability:
\BE\nabla_{\bf X}\times(\At\LA{\bf h}_0\RA)=\lambda_1\LA{\bf h}_0\RA.\EE{av1}
Here $\At$ denotes the tensor of magnetic $\alpha$-effect, a $3\times3$
matrix whose columns are
\BE\At_k=\LA{\bf v}\times{\bf s}_k\RA.\EE{at}
It is independent of the spatial and temporal variables; this significantly
simplifies the study of the spectrum of the so-called $\alpha$-effect operator
encountered in the l.h.s.~of \rf{av1} (see section \rf{spec}). The entries
of the $\alpha$-effect tensor are denoted $\A^m_k$.

Evidently, for $\lambda_1\ne0$, \rf{so} holds true for $n=0$ automatically.

\section{The oscillogenic $\alpha$-effect for a flow antisymmetric
in a Cartesian variable}\label{osc}

In this section we show that, as a consequence of the assumed mirror
antisymmetry of the flow about the plane $x_1=0$ (defined by \rf{as}
for $j=1$), the magnetic $\alpha$-effect tensor $\At$ \rf{at} has a peculiar
structure combining matrix symmetry and antisymmetry: its lower right
$2\times2$ submatrix is antisymmetric, the entire main diagonal is populated
with zeroes, but the left column is symmetric to the upper row. Having
established these properties in section \ref{stru}, in section \ref{spec}
we demonstrate that if the main field is assumed to be periodic in the slow
spatial variables, the periodicity cell being the cube $\T^3$, then
the spectrum of the $\alpha$-effect operator consists of imaginary numbers,
and thus the $\alpha$-effect in such a flow is oscillogenic.

Our arguments will be based on the formulae presented on p.~34
of \cite{VZ} which relate the $\alpha$-effect to the helicities\footnote{Note
that this is not the magnetic helicity that is defined as the mean scalar
product of magnetic field and its vector potential.} and the cross-helicities
of the electric current densities $\nabla\times{\bf s}_k$ associated
(by the Maxwell--Amp\`ere law) with the magnetic fields ${\bf s}_k$.
For reader's convenience, we now derive them. ``Uncurling'' of the eigenvalue
equation \rf{Seq} yields
\BE-\eta\,\nabla\times{\bf s}_k+{\bf v}\times{\bf s}_k=
\LA{\bf v}\times{\bf s}_k\RA+\nabla p_k,\EE{unc}
where $p_k({\bf x})$ are suitable space-periodic functions. Scalar multiplying
this relation by ${\bf s}_m$ and averaging the product over $\T^3$ we find
$$-\eta\,\LA{\bf s}_m\cdot\nabla\times{\bf s}_k\RA
+\LA{\bf s}_m\cdot({\bf v}\times{\bf s}_k)\RA=\A^m_k,$$
whereby
\BE-2\eta\,\LA{\bf s}_m\cdot\nabla\times{\bf s}_k\RA=\A^m_k+\A^k_m\EE{Amk}
(we have used the self-adjointness of the curl), and for $k=m$
\BE-\eta\,\LA{\bf s}_k\cdot\nabla\times{\bf s}_k\RA=\A^k_k.\EE{Akk}
(More precisely, we have now linked the symmetric part of the $\alpha$-effect
tensor, $(\A^m_k+\A^k_m)/2$, to the current helicities and cross-helicities
${\bf s}_m\cdot\nabla\times{\bf s}_k$. However, the antisymmetric part
of the tensor, $(\A^m_k-\A^k_m)/2$, controls only imaginary parts
of eigenvalues of the $\alpha$-effect operator \cite{RCZ}, and thus the growth
rates due to the action of the $\alpha$-effect are fully determined
by the current helicities and cross-helicities under discussion.)

\subsection{The structure of the $\alpha$-effect tensor}\label{stru}
We consider henceforth flows $\bf v$ that are antisymmetric in one variable,
say, $x_1$. We show that in this case all diagonal entries $\A^k_k$ are
zero, and the non-diagonal entries of $\At$ are linked by certain relations
(see \rf{snd} and \rf{and} below).
Note that the curl transforms fields symmetric in $x_1$ into antisymmetric
ones and vice versa, and for $\bf v$ antisymmetric in $x_1$, vector
multiplication by $\bf v$ of a vector field preserves its symmetry or
antisymmetry in $x_1$.
We decompose the fields ${\bf s}_k$ into symmetric and antisymmetric parts,
\BE\ss_k({\bf x})={1\over2}\left(\left[\begin{array}{r}
-{\bf s}^1_k(-x_1,x_2,x_3)\\{\bf s}^2_k(-x_1,x_2,x_3)\\{\bf s}^3_k(-x_1,x_2,x_3)\\
\end{array}\right]+{\bf s}_k({\bf x})\right),\qquad\sa_k={\bf s}_k-\ss_k,\EE{ssa}
respectively. By virtue of \rf{unc} and the symmetry properties mentioned above,
\be-\eta\,\nabla\times\ss_k+{\bf v}\times\sa_k&=\Aa_k+\nabla p^a_k,\label{ssk}\\
-\eta\,\nabla\times\sa_k+{\bf v}\times\ss_k&=\As_k+\nabla p^s_k,\label{sak}\\
\LA\ss_1\RA=0,\quad\LA\sa_1\RA={\bf e}_1;\qquad\LA\ss_k\RA&={\bf e}_k,\quad
\LA\sa_k\RA=0\mbox{ \ for \ }k=2,3.\label{ave}\end{align}
Here constant vectors
$$\As_k=\left[\begin{array}{c}0\\\A^2_k\\\A^3_k\end{array}\right],
\qquad\Aa_k=\left[\begin{array}{c}\A^1_k\\0\\0\end{array}\right]$$
are the symmetric and antisymmetric part of the mean vector
$\LA{\bf v}\times{\bf s}_k\RA$, and space-periodic scalar functions $p^a_k$
and $p^s_k$ are the odd in $x_1$ and even part of $p_k$, respectively.

Since the curl is a self-adjoint operator, \rf{Akk} reduces to
$$\A^k_k=-2\eta\,\LA\ss_k\cdot\nabla\times\sa_k\RA
=-2\eta\,\LA\sa_k\cdot\nabla\times\ss_k\RA.$$
For $k=1$, we scalar multiply \rf{sak} by $\ss_1$, average over $\T^3$,
use \rf{ave} and find $\LA\ss_1\cdot\nabla\times\sa_1\RA=0$.\break
For $k>1$, scalar multiplication of \rf{ssk} by $\sa_k$ followed by the
same transformations yields\break$\LA\sa_k\cdot\nabla\times\ss_k\RA=0$. Thus,
for all $k$,
\BE\A^k_k=0.\EE{zero}

Relations between non-diagonal entries of the $\alpha$-tensor can be derived
as follows. In terms of symmetric and antisymmetric parts of the neutral modes,
\rf{Amk} becomes
\BE-2\eta\,\LA\sa_m\cdot\nabla\times\ss_k
+\sa_k\cdot\nabla\times\ss_m\RA=\A^m_k+\A^k_m\EE{Smk}
for all $m$ and $k$. Suppose $k>1$ and $m>1$. We scalar multiply \rf{ssk}
by $\sa_m$, average over $\T^3$, use \rf{ave}, symmetrise in $m$ and $k$,
and find that the l.h.s.~of \rf{Smk} is zero. Therefore,
\BE\A^3_2=-\A^2_3.\EE{snd}

To derive two remaining identities for entries of $\At$, we scalar multiply
\rf{sak} by $\ss_m$, average over $\T^3$ and symmetrise in $m$ and $k$.
Letting now $k=1$ and $m>1$, we use \rf{ave} to find
$$-\eta\,\LA\ss_m\cdot\nabla\times\sa_1+\ss_1\cdot\nabla\times\sa_m\RA=\A^m_1.$$
By comparison with \rf{Smk},
\BE\A^m_1=\A^1_m\mbox{ \ for \ }m=2,3.\EE{and}

\subsection{The spectrum of the $\alpha$-effect operator}\label{spec}
Let us consider the eigenvalue equation \rf{av1} for the $\alpha$-effect operator
\BE\nabla_{\bf X}\times(\At\LA{\bf h}_0\RA)=\lambda_1\LA{\bf h}_0\RA,\qquad
\nabla_{\bf X}\cdot\LA{\bf h}_0\RA=0.\EE{aleq}
Proceeding as in \cite{RCZ} (where the solution to \rf{aleq} was derived
for an arbitrary matrix $\At$), we assume that the mean magnetic field
is space-periodic and hence eigenfunctions are Fourier harmonics:
\BE\LA{\bf h}_0\RA={\bf H}\e.\EE{fou}
Here $\bf q$ and $\bf H$ are constant vectors. (On the one hand, this choice
is natural, since the initial large-scale magnetic fields residing
in the entire space can be expanded in the Fourier series, if it is
space-periodic, or Fourier integral otherwise, implying that the respective
time-dependent solution to the two-scale kinematic dynamo problem is a linear
combination of the solutions considered here; we will thus study
the temporal behaviour of building blocks for such expansions.
On the other, the approach can be generalised by considering finite,
in the slow spatial variables, volumes of fluid and setting appropriate
boundary conditions for the mean magnetic field; the eigenfunctions
of the $\alpha$-effect operator will then have a different structure.)
Further assuming that the wave vector is unit, $|{\bf q}|=1$, we express it
in the spherical coordinates whose axis is aligned with the $x_1$-axis:
\BE q_1=\cos\theta,\qquad q_2=\sin\theta\cos\varphi,\qquad q_3=\sin\theta\sin\varphi.\EE{wv}
The solenoidality of $\LA{\bf h}_0\RA$ (see \rf{so}) is then equivalent
to the orthogonality relation
$${\bf H\cdot q}=0$$
implying
\BE{\bf H}=\Theta_{\rm t}{\bf q}^{\rm t}+\Theta_{\rm p}{\bf q}^{\rm p}.\EE{Hpt}
Here we have introduced vectors
\BE{\bf q}^{\rm t}=(0,-\sin\varphi,\cos\varphi),\qquad
{\bf q}^{\rm p}=(-\sin\theta,\cos\theta\cos\varphi,\cos\theta\sin\varphi)\EE{qpt}
that constitute, together with $\bf q$, an orthonormal basis of positive
orientation in $\R^3$. We substitute \rf{fou} into \rf{aleq} and scalar multiply
the resultant equation by ${\bf q}^{\rm t}$ and ${\bf q}^{\rm p}$, which
transforms \rf{aleq} into an equivalent eigenvalue problem for a $2\times2$
matrix:
\BE\i\left[\begin{array}{rr}
{\bf q}^{\rm p}\cdot\At{\bf q}^{\rm t}&{\bf q}^{\rm p}\cdot\At{\bf q}^{\rm p}\\
-{\bf q}^{\rm t}\cdot\At{\bf q}^{\rm t}&-{\bf q}^{\rm t}\cdot\At{\bf q}^{\rm p}\end{array}\right]
{\bf\Theta}=\lambda_1{\bf\Theta},\EE{ei}
where ${\bf\Theta}=\left[\begin{array}{l}\Theta_{\rm t}\\ \Theta_{\rm p}\end{array}\right]$.

As shown in the previous section, for a flow antisymmetric in $x_1$,
the entries of the $\alpha$-effect tensor satisfy relations \rf{zero},
\rf{snd} and \rf{and}. Consequently, the matrix in the l.h.s.~of \rf{ei} is
\BE\left[\begin{array}{cc}
\A^2_3\cos\theta-(\A^3_1\cos\varphi-\A^1_2\sin\varphi)\sin\theta
&-(\A^3_1\sin\varphi+\A^1_2\cos\varphi)\sin\,2\theta\\
0&\A^2_3\cos\theta+(\A^3_1\cos\varphi-\A^1_2\sin\varphi)\sin\theta
\end{array}\right].\EE{A2}
As a result, both eigenvalues of problem \rf{ei} are imaginary:
\se{l1}\be\lambda_1^\pm&=\i\,(\A^2_3\cos\theta\pm(\A^3_1\cos\varphi
-\A^1_2\sin\varphi)\sin\theta)=\i\,(\A^2_3q_1\pm(\A^3_1q_2-\A^1_2q_3)),\label{lpm}\\
{\bf\Theta}^+&=\left[\begin{array}{c}-(\A^3_1\sin\varphi+\A^1_2\cos\varphi)\cos\theta\\
\A^3_1\cos\varphi-\A^1_2\sin\varphi\end{array}\right],\qquad
{\bf\Theta}^-=\left[\begin{array}{c}1\\0\end{array}\right],\label{Tpm}\\
{\bf H}^+&=(-(\A^3_1\cos\varphi-\A^1_2\sin\varphi)\sin\theta,\,
\A^3_1\cos\theta,\,-\A^1_2\cos\theta),\qquad
{\bf H}^-=(0,-\sin\varphi,\cos\varphi).\label{Hpm}\end{align}\end{subequations}
Thus, any flow antisymmetric in $x_1$ features
the oscillogenic $\alpha$-effect. Depending on the wave vector $\bf q$,
the frequency of oscillations in the slow time $T_1$ varies
between zero and $(\A^2_{32}+\A^2_{13}+\A^2_{21})^{1/2}$.

\section{Magnetic eddy diffusivity}\label{hi}

We continue now to study equations for $n=1$ and 2
from the hierarchy emerging from \rf{hie}. The solvability condition for
the order $\varepsilon^2$ equation (see section \ref{o2}) reveals that
magnetic eddy diffusivity action on the mean field $\LA{\bf h}_0\RA$
is described by two tensors, $\Dw$ and $\Db$. They are expressed in terms
of solutions to auxiliary problems for the operator $\L^*$ \rf{La}, adjoint
to the operator
of magnetic induction, and important conclusions about the solutions based on
the mirror antisymmetry of the flow are drawn in section \ref{Zs}.
This enables us to determine in the two subsequent sections the structure
of the two tensors, which combine symmetry and antisymmetry properties
(similar to those of the $\alpha$-effect tensor $\At$). This significantly
simplifies expressions for the growth rate of the magnetic mode in the
slow time $T_2=\varepsilon t^2$ (see section \ref{gr}). The growth rates
have singular behaviour in the azimuthal direction $\varphi$ of the wave vector
of the mode, guaranteeing the action of the large-scale dynamo for whichever
large molecular diffusivity provided the scale ratio $\varepsilon$ is
sufficiently small. In section \ref{le} we consider the power series expansion
of the eddy diffusivity tensors and make estimations, how close the
azimuthal direction $\varphi$ of the wave vector must be to the singular
value for making possible the large-scale generation by the mechanism of
negative eddy diffusivity.

\subsection{Order $\varepsilon^1$ equation, continued}
Substituting \rf{fou} into the fluctuating part of \rf{eq_1} yields
$$\L{\bf h}_1+\e\left(2\i\eta\sum_{k=1}^3H_k
({\bf q}\cdot\nabla){\bf s}_k+\i{\bf q}\times\left\{{\bf v}\times
\sum_{k=1}^3H_k{\bf s}_k\right\}-\lambda_1\sum_{k=1}^3H_k\{{\bf s}_k\}\right)=0.$$
The solvability condition \rf{av1} for this equation is satisfied.
By linearity of the small-scale operator~$\L$,
\BE{\bf h}_1=\sum_{k=1}^3\left(\LA{\bf h}_1\RA_k{\bf s}_k+\e H_k\left(
\lambda_1{\bm\gamma}_k+\i\sum_{m=1}^3q_m{\bf g}_{mk}\right)\right),\EE{sol_1}
where ${\bm\gamma}_k({\bf x})$ and ${\bf g}_{mk}({\bf x})$ are small-scale
zero-mean space-periodic solutions to {\it auxiliary problems of types II
and II$\,'$}, respectively:
\BE\L{\bf g}_{mk}=-2\eta{\partial{\bf s}_k\over\partial x_m}
-{\bf e}_m\times\{{\bf v}\times{\bf s}_k\},\EE{II}
\BE\L{\bm\gamma}_k=\{{\bf s}_k\}.\EE{aII}

\subsection{Order $\varepsilon^2$ equation}\label{o2}
We infer from \rf{hie}, for $n=2$,
$$\L{\bf h}_2+2\eta(\nabla\cdot\nabla_{\bf X}){\bf h}_1
+\eta\nabla_{\bf X}^2{\bf h}_0+\nabla_{\bf X}\times({\bf v}\times{\bf h}_1)
=\lambda_1{\bf h}_1+\lambda_2{\bf h}_0$$
and derive the solvability condition for this equation by averaging it
in the fast variables and substituting \rf{fou} and \rf{sol_1}:
\BE\nabla_{\bf X}\times(\At\LA{\bf h}_1\RA)+\i\e{\bf q}\times\sum_{k=1}^3
H_k\left(\lambda_1\Dw_k+\i\sum_{m=1}^3q_m\Db_{mk}\right)
=\lambda_1\LA{\bf h}_1\RA+(\lambda_2+\eta)\e{\bf H}.\EE{av2}
Here we have denoted
\BE\Dw_k=\LA{\bf v}\times{\bm\gamma}_k\RA,\qquad
\Db_{mk}=\LA{\bf v}\times{\bf g}_{mk}\RA.\EE{Dmk}
In \rf{av2}, both terms independent of
$\LA{\bf h}_1\RA$ are proportional to $\e$. Consequently,
$$\LA{\bf h}_1\RA=\e{\bf H}',$$
where ${\bf H}'$ satisfies the equation
\BE\i{\bf q}\times\sum_{k=1}^3H_k\left(\lambda_1\Dw_k+\i\sum_{m=1}^3q_m\Db_{mk}
\right)=-\i{\bf q}\times(\At{\bf H}')+\lambda_1{\bf H}'+(\lambda_2+\eta){\bf H}.\EE{h_eq}

In what follows we assume $\lambda_1^+\ne\lambda_1^-$, implying that ${\bf H}^+$
and ${\bf H}^-$ constitute a basis in the subspace of three-dimensional vectors
orthogonal to $\bf q$, and focus on an eigensolution \rf{lpm} of problem
\rf{ei}, $\lambda_1=\lambda_1^\sigma$ ($\sigma$ denoting $+$ or $-$) and
the associated vector ${\bf H}^\sigma$ \rf{Hpm}. Let $\sigma_2$ denote the sign
opposite to $\sigma$; we will mark by superscripts $\sigma$ and $\sigma_2$
the quantities pertaining to the respective sign in \rf{l1}.
The solenoidality condition \rf{so} for $n=1$ implies an expansion
${\bf H}'=\beta^\sigma{\bf H}^\sigma+\beta^\sigma_2{\bf H}^{\sigma_2}$.
By virtue of \rf{aleq}, \rf{h_eq} takes the form
\BE{\bf q}\times\sum_{k=1}^3H^\sigma_k\left(\i\lambda^\sigma_1\Dw_k
-\sum_{m=1}^3q_m\Db_{mk}\right)
=\beta^\sigma_2(\lambda_1^\sigma-\lambda_1^{\sigma_2}){\bf H}^{\sigma_2}
+(\lambda_2^\sigma+\eta){\bf H}^\sigma.\EE{bl2}
Note that the coefficient $\beta^\sigma$ does not enter \rf{bl2}.
Indeed, an eigenfunction $\bf h$ of the dynamo problem \rf{Main} can only
be determined up to a constant (in the spatial variables) factor;
multiplying $\bf h$ by linear functions in $\varepsilon$ arbitrarily alters
$\beta^\sigma$. A normalisation condition $\beta^\sigma=0$ can be prescribed.

We form the triple product of \rf{bl2} with $\bf q$ and ${\bf H}^{\sigma_2}$,
use the orthogonality ${\bf H}^{\sigma_2}\cdot{\bf q}=0$ and the fact that
$\{\bf q,q^{\rm p},q^{\rm t}\}$ is an orthonormal basis of positive
orientation in $\R^3$; this yields
\BE\lambda^\sigma_2={\sum_{k=1}^3H^\sigma_k\left(\i\lambda_1^\sigma
\Dw_k-\sum_{m=1}^3q_m\Db_{mk}\right)\cdot{\bf H}^{\sigma_2}\over
\Theta^{\sigma_2}_p\Theta^\sigma_t-\Theta^{\sigma_2}_t\Theta^\sigma_p}-\eta,\EE{l2}
whereby $\lambda_2^\pm$ are real. The same procedure with the use
of ${\bf H}^\sigma$ instead of ${\bf H}^{\sigma_2}$ yields
$$\beta^\sigma_2={\sum_{k=1}^3H^\sigma_k\left(\i\lambda_1^\sigma\Dw_k
-\sum_{m=1}^3q_m\Db_{mk}\right)
\cdot{\bf H}^\sigma\over(\lambda_1^{\sigma_2}-\lambda_1^\sigma)
(\Theta^{\sigma_2}_p\Theta^\sigma_t-\Theta^{\sigma_2}_t\Theta^\sigma_p)}.$$

\subsection{Consequences of the antisymmetry in $x_1$ of the flow}\label{Zs}
As usual, it is convenient to consider {\it auxiliary problems for the adjoint
operator} \cite{VZ}:
\BE\L^*{\bf Z}_l={\bf v}\times{\bf e}_l-\LA{\bf v}\times{\rm^r}{\bf s}_l\RA,\EE{adj}
whose solutions ${\bf Z}_l$ are assumed to be zero-mean; the adjoint operator
$\L^*$ is defined by \rf{La}. Here and in what follows, the superscript ``r"
marks objects pertinent to the reverse flow $-\bf v(x)$:
\begin{align*}{\rm^r\!}\L{\bf h}=\eta\nabla^2{\bf h}-\nabla
\times({\bf v}\times{\bf h}),\qquad&{\rm^r\!}\L\,{\rm^r}{\bf s}_k=0,\\
{\rm^r\!}\L^*{\bf h}=\eta\nabla^2{\bf h}+{\bf v}\times(\nabla\times{\bf h}),
\qquad&{\rm^r\!}\L^*\,\rZ_l=-{\bf v}\times{\bf e}_l
+\LA{\bf v}\times{\bf s}_l\RA.\end{align*}

In view of \rf{adj}, \rf{Dmk} and \rf{aII},
\BE\Dt^l_k=-\LA\L^*{\bf Z}_l\cdot{\bm\gamma}_k\RA
=-\LA{\bf Z}_l\cdot{\bf s}_k\RA;\EE{Dlk}
similarly, by virtue of \rf{adj}, \rf{Dmk} and \rf{II},
\BE\D^l_{mk}=-\LA\L^*{\bf Z}_l\cdot{\bf g}_{mk}\RA
=\LA{\bf Z}_l\cdot\left(2\eta{\partial{\bf s}_k\over\partial x_m}
+{\bf e}_m\times({\bf v}\times{\bf s}_k)\right)\RA.\EE{Dlmk}

We will need an expression for the entries $\D^l_{mk}$ in terms of solutions
${\bf Z}_l$ and $\rZ_l$ to auxiliary problems for the adjoint
operator for the direct, $\bf v(x)$, and reverse, $-\bf v(x)$, flows,
respectively. Clearly, \rf{adj} implies
${\rm^r\!}\L(\nabla\times{\bf Z}_l+{\bf e}_l)=0.$
Therefore, for the generic data $\eta$ and $\bf v(x)$,
\se{ZvS}\BE\nabla\times{\bf Z}_l+{\bf e}_l={\rm^r}{\bf s}_l\quad\Rightarrow\quad
{\bf Z}_l=\eta^{-1}\nabla^{-2}\{{\bf v}\times{\rm^r}{\bf s}_l\},\EE{ZS}
where $\nabla^{-2}$ denotes the inverse Laplacian in the fast variables.
Similarly,
\BE\nabla\times\rZ_k+{\bf e}_k={\bf s}_k\quad\Rightarrow\quad
\rZ_k=-\eta^{-1}\nabla^{-2}\{{\bf v}\times{\bf s}_k\}.\EE{ZmS}\end{subequations}

Actually, existence of the solution \rf{ZS} to problem \rf{adj} has
an implication for the entries of the $\alpha$-tensors for the direct and
reverse flows, $\A^l_k=\LA{\bf v}\times{\bf s}_k\RA_l$ and
${\rm^r}\A^k_l=-\LA{\bf v}\times{\rm^r}{\bf s}_l\RA_k$, respectively: the solvability
condition for \rf{adj} consists of the orthogonality of the r.h.s.~of this
equation to the kernel of the operator adjoint to $\L^*$, i.e., to all
the three ${\bf s}_k({\bf x})$, whereby
$$0=\LA({\bf v}\times{\bf e}_l-\LA{\bf v}\times{\rm^r}{\bf s}_l\RA)\cdot{\bf s}_k\RA
=-\A^l_k+{\rm^r}\A^k_l,$$
i.e., $\A^l_k={\rm^r}\A^k_l$ for all $l$ and $k$. This identity was proven
by a different argument in \cite{RCZ}.

While so far the presentation in this subsection has not relied on any
symmetry or antisymmetry of the flow,
in the remainder we consider flows antisymmetric in $x_1$.
Let us decompose the fields ${\bf s}_k$ and ${\bf Z}_k$ into symmetric and
antisymmetric parts (see \rf{ssa}) which we will mark by the superscripts
``s" and ``a", respectively. Since the curl transforms fields symmetric in $x_1$
into antisymmetric ones and vice versa, and the Laplacian as well as vector
multiplication by $\bf v$ preserves the symmetry or antisymmetry in $x_1$,
the symmetric and antisymmetric parts of \rf{Seq} state:
$$\eta\nabla^2\ss_k+\nabla\times({\bf v}\times\sa_k)=0,\qquad
\eta\nabla^2\sa_k+\nabla\times({\bf v}\times\ss_k)=0.$$
These relations imply
\BE{\rm^r}{\bf s}_k=o_k(\ss_k-\sa_k),\EE{xsa}
where $o_k=-1$ for $k=1$ and $o_k=1$ for $k=2,3$ so that the condition
$\LA{\rm^r}{\bf s}_k\RA={\bf e}_k$ is satisfied.
Consequently, we find from \rf{ZvS}
\BE{\bf Z}_l=\Zs_l+\Za_l,\qquad\rZ_k=o_k(\Za_k-\Zs_k).\EE{Zas}
Thus, for evaluation of the entries of tensors \rf{Dmk} for a flow
antisymmetric in a Cartesian variable ($x_1$ in our case) it suffices
to solve the three auxiliary problems of type I and then to use \rf{Dlk},
\rf{Dlmk}, \rf{ZS} and \rf{xsa}.

\subsection{The structure of the tensor $\Dw$}\label{sD2}
Let us establish relations for the entries of tensor $\Dw$ \rf{Dmk} involved
in the homogenised magnetic induction operator.
Using the solenoidality of ${\bf s}_k$ and ${\rm^r}{\bf s}_l$,
the self-adjointness of the curl and \rf{ZS}, we transform \rf{Dlk}:
$$\Dt^l_k=\LA(\nabla\times{\bf Z}_l)\cdot\nabla^{-2}\nabla\times{\bf s}_k\RA
=\LA{\rm^r}{\bf s}_l\cdot\nabla^{-2}\nabla\times{\bf s}_k\RA.$$
By virtue of \rf{xsa} and since the curl maps an antisymmetric
field into a symmetric one and vice versa,
$$\Dt^l_k=o_l\LA\ss_l\cdot\nabla^{-2}\nabla\times
\sa_k-\sa_l\cdot\nabla^{-2}\nabla\times\ss_k\RA.$$
Since the curl and the Laplacian are self-adjoint, this expression implies
\BE\Dt^l_k=-o_lo_k\Dt^k_l\qquad\Rightarrow\qquad\Dt^k_k=0~~\mbox{for all}~k,
\quad\Dt^2_3=-\Dt^3_2,\quad\Dt^1_2=\Dt^2_1,\quad\Dt^1_3=\Dt^3_1,\EE{Dti}
which mimicks the properties of the $\alpha$-tensor \rf{zero}, \rf{snd} and
\rf{and}.

\subsection{The structure of the tensor $\Db$}\label{sD3}
Using relations \rf{ZmS} to eliminate ${\bf s}_k$ in \rf{Dlmk},
we express $\D^l_{mk}$ as a bilinear form of solutions to auxiliary problems
for the adjoint operator \cite{ABNZ},
\se{dt}\BE\D^l_{mk}=\B_m({\bf Z}_l,\rZ_k),\EE{ZlZk}
where
\BE\B_m({\bf F,H})=\eta\LA{\bf F}\cdot\left(2\,\nabla\times{\partial{\bf H}
\over\partial x_m}-{\bf e}_m\times\nabla^2{\bf H}\right)\RA.\EE{Bm}\end{subequations}
Since the Laplacian and the curl are self-adjoint operators, and the triple
product is antisymmetric with respect to permutation of its factors,
for all vector fields $\bf F$ and $\bf H$
\BE\B_m({\bf F,H})=-\B_m({\bf H,F})\quad\Rightarrow\quad\B_m({\bf F,F})=0.\EE{aB}

We now consider implications of the antisymmetry of the flow in $x_1$
for the structure of the magnetic eddy diffusivity tensor \rf{dt}.
Substituting relations \rf{Zas} into \rf{ZlZk} and using the antisymmetry
\rf{aB} of the bilinear form $\B_m$, we derive 12 identities:\\
$i$. For all $k$,
\se{Di}\BE\D^k_{1k}=2o_k\B_1(\Zs_k,\Za_k)=0,\EE{Di1}
since for $m=1$ the operator acting in \rf{Bm} on the second argument
of the bilinear form
$\B_m$ preserves the symmetry or the antisymmetry of this argument.\\
$ii$. For the same reasons,
$$\D^l_{1k}=o_k\B_1(\Zs_l+\Za_l,\Za_k-\Zs_k)
=o_k(\B_1(\Za_l,\Za_k)-\B_1(\Zs_l,\Zs_k))=-o_k\B_1(\Zs_k+\Za_k,\Za_l-\Zs_l)
=-o_lo_k\D^k_{1l}$$
\BE\Rightarrow\qquad\D^2_{11}=\D^1_{12},\quad\D^3_{11}=\D^1_{13}\mbox{~~and~~}\D^2_{13}=-\D^3_{12}.\EE{Di2}
$iii$. For $m=2,3$, the second factor in the scalar product defining $\B_m$ is
a symmetric vector field when the second argument of the form is antisymmetric,
and an antisymmetric one when the argument is symmetric. Consequently,
$$\D^l_{mk}=o_k\B_m(\Zs_l+\Za_l,\Za_k-\Zs_k)
=o_k(\B_m(\Zs_l,\Za_k)-\B_m(\Za_l,\Zs_k))=o_k\B_m(\Zs_k+\Za_k,\Za_l-\Zs_l)
=o_lo_k\D^k_{ml}$$
\BE\Rightarrow\quad\D^2_{21}=-\D^1_{22},\ \ \D^3_{21}=-\D^1_{23},
\ \ \D^2_{31}=-\D^1_{32},\ \ \D^3_{31}=-\D^1_{33},\ \ \D^3_{22}=\D^2_{23}
\mbox{~~and~~}\D^3_{32}=\D^2_{33}.\EE{Di3}\end{subequations}

\subsection{The growth rate}\label{gr}
In view of the 6 identities \rf{Dti} for tensor $\Dw$ and the 12 identities
\rf{Di} for $\Db$, \rf{l2} implies
\se{rpm}
\BE\lambda_2^\pm=Q_1+Q_2+(Q_1-Q_2)\cos2\theta\pm Q_3\sin2\theta-\eta,\EE{q0}
where
\be Q_1=&\,-{1\over2}\,(\Dt^2_3\A^2_3+\D_{13}^2),\label{q1}\\
Q_2=&\,{1\over4}\Big((\D_{23}^1+\D_{32}^1+\A^1_2\Dt^1_2-\A^3_1\Dt^3_1)\cos2\varphi
+(\D_{33}^1-\D_{22}^1+\A^3_1\Dt^1_2+\A^1_2\Dt^3_1)\sin2\varphi\nonumber\\
&+\,\D_{23}^1-\D_{32}^1-\A^1_2\Dt^1_2-\A^3_1\Dt^3_1\Big),\label{q2}\\
Q_3=&\,{\cos\varphi\over2}(\D_{23}^2+\D_{32}^2-\D_{13}^1-\A^3_1\Dt^2_3-\A^2_3\Dt^3_1)
-{\sin\varphi\over2}(\D_{23}^3+\D_{32}^3-\D_{12}^1-\A^2_3\Dt^1_2-\A^1_2\Dt^2_3)\nonumber\\
&+{2\A^1_2\D_{23}^3+2\A^3_1\D_{32}^2-\Big(\A^3_1(2\D_{32}^3+\D_{23}^3-\D_{22}^2)
+\A^1_2(2\D_{23}^2+\D_{32}^2-\D_{33}^3)\Big)\sin2\varphi\over
4(\A^1_2\sin\varphi-\A^3_1\cos\varphi)}.\label{q3}\end{align}\end{subequations}
It is evident from these expressions that $\lambda_2$ (as well as $\lambda_1$
\rf{lpm}) does not depend on $\varphi$ when $\sin\theta=0$ (this is just
a condition for the geometric consistency of expression \rf{q0} for the eigenvalue).
For a fixed $\varphi$, the maximum over $\theta$ of growth rate \rf{q0},
\BE\max_{0\le\theta\le\pi}\lambda_2^\pm=Q_1+Q_2+((Q_1-Q_2)^2+Q_3^2)^{1/2}-\eta,\EE{grm}
is obtained when $\tan2\theta=Q_3/(Q_1-Q_2)$. Minimum eddy diffusivity
is a function of the azimuthal direction:
$\eta_{\rm eddy}(\varphi)=-\max_{0\le\theta\le\pi}\lambda_2^\pm$. Generically,
for any $\theta$ that is not an integer multiple of $\pi/2$, eddy diffusivity
is guaranteed to be negative for some $\varphi$: $\lambda_2^\pm$
do take both positive and negative values on varying $\varphi$, because
for any integer $n$ the denominator in $Q_3$ changes the sign at
\BE\fs=\arctan(\A^3_1/\A^1_2)+n\pi,\EE{pas}
resulting in a singularity in $Q_3$
unless for $\varphi=\fs$ the numerator in $Q_3$ also vanishes.

Clearly, $Q_1$ and $Q_2$ are $\pi$-periodic in $\varphi$, and $Q_3$ only changes
the sign when $\varphi$ increases by $\pi$, a half of the period. Consequently,
\rf{rpm} yields $\lambda_2^+(\theta,\varphi)=\lambda_2^-(\theta,\varphi+\pi)$
and \rf{grm} implies the $\pi$-periodicity of $\max_{0\le\theta\le\pi}\lambda_2^\pm$.
Also note that both $\lambda_2^\pm$ are invariant under the mapping
$\varphi\mapsto\varphi+\pi$, $\theta\mapsto\pi-\theta$.

Of course, the presence of the singularity in \rf{q3} does not imply that
the dynamo under consideration actually features infinitely large growth rates
\rf{e} in the fast time $t$;
rather, it just signals that at the point of singularity the asymptotics ansatz
\rf{ps} breaks down. Arbitrarily large growth rates in the slow time
scale $T_2=\varepsilon^2 t$ can indeed be realised, but this requires
to sufficiently decrease the scale ratio $\varepsilon$, so that the leading term
of the asymptotics is not offset by the subdominant terms of the expansion.

\subsection{Large $\eta$ asymptotics}\label{le}
Since for whichever large molecular diffusivity $\eta$ eddy diffusivity
$\eta_{\rm eddy}(\varphi)$ takes negative values, it makes sense to investigate
the $\alpha$-effect and eddy diffusivity tensors in the limit $\eta\to\infty$.
Here this limit is considered.

For large $\eta$, neutral modes can be expanded in power series in $\eta^{-1}$:
\BE{\bf s}_k({\bf x})=\sum_{j=0}^\infty{\bf s}_{k,j}({\bf x})\eta^{-j}.\EE{sk}
For $j>0$, the solenoidal zero-mean coefficients satisfy the recurrence relations
\BE{\bf s}_{k,j}({\bf x})=-\nabla\times\nabla^{-2}({\bf v}\times
{\bf s}_{k,j-1}),\qquad{\bf s}_{k,0}={\bf e}_k;\EE{re}
in particular, ${\bf s}_{k,1}=-\nabla^{-2}{\partial{\bf v}/\partial x_k}$.
By \rf{re}, for a sufficiently regular flow $\bf v$ the operator that yields
${\bf s}_{k,j}$
from ${\bf s}_{k,j-1}$ is bounded in the Sobolev space $H^1(\T^3)$:
$$\|\nabla\times\nabla^{-2}({\bf v}\times{\bf f})\|\le C\|{\bf f}\|,$$
where $\|\cdot\|$ denotes the norm in $H^1(\T^3)$ and constant $C$ is
independent of an arbitrary field $\bf f$ from $H^1(\T^3)$. (Using H\"older
inequality, it is easy to show that $C\le S|{\bf v}|_3$, where $|\cdot|_p$
denotes the norm in the Lebesgue space L$_p(\T^3)$ and $S$ is a constant
in the inequality $|{\bf f}|_6\le S\|{\bf f}\|$ following from the Sobolev
embedding theorem for space-periodic zero-mean fields.) Therefore, series
\rf{sk} is majorised by a geometric series
with the ratio $C/\eta$ and is guaranteed to converge for $\eta>C$.
Numerical algorithms for computation of the $\alpha$-effect and eddy
diffusivity tensors based on the expansion \rf{sk} and employing Pad\'e
approximation will be considered in \cite{GCZ}.

Relation \rf{ZS} implies an expansion
$${\bf Z}_k=\sum_{j=1}^\infty{\bf Z}_{k,j}({\bf x})\eta^{-j}.$$
By comparison of recurrence relations \rf{re} for the direct and reverse flow,
coefficients of the respective fields for the reverse flow (also marked
by the superscript ``r") satisfy
\BE{\rm^r}{\bf s}_{k,j}({\bf x})=(-1)^j{\bf s}_{k,j}({\bf x}),\qquad
\rZ_{k,j}=(-1)^j{\bf Z}_{k,j}({\bf x}),\EE{rZc}
and thus
\BE{\bf Z}_{k,j}=\,(-1)^{j-1}\nabla^{-2}({\bf v}\times{\bf s}_{k,j-1}).\EE{Za}
Consequently, the tensors defined by relations \rf{at}, \rf{Dlk} and \rf{dt}
are also expandable in power series:
$$\At_k=\sum_{j=1}^\infty\At_{k,j}\eta^{-j},\qquad
\Db_{mk}=\sum_{j=1}^\infty\Db_{mk,j}\eta^{-j},\qquad
\Dw_k=\sum_{j=1}^\infty\Dw_{k,j}\eta^{-j}.$$
The coefficients in the series for the $\alpha$-effect tensor are
$\At_{k,j}=\LA{\bf v}\times{\bf s}_{k,j}\RA$, in particular,
\BE\At_{k,1}=-\LA{\bf v}\times\nabla^{-2}{\partial{\bf v}\over\partial x_k}\RA.\EE{ata}
The leading term in the series for $\Dw$ has the coefficient
\BE\Dw_{k,2}=-\LA\nabla^{-2}{\bf v}\times
\nabla^{-2}{\partial{\bf v}\over\partial x_k}\RA.\EE{Dta}

Now, by \rf{dt},
\BE\D^l_{mk,1}=\LA(\nabla^{-2}{\bf v}\times{\bf e}_l)\cdot
\left(-2{\partial^2\over\partial x_m\partial x_k}\nabla^{-2}{\bf v}
+{\bf e}_m\times({\bf v}\times{\bf e}_k)\right)\RA
=\epsilon_{nkl}\LA v^m\nabla^{-2}v^n\RA,\EE{D1}
where\footnote{Summation over repeated indices is not tacitly assumed.}
$n=6-l-k$ for $l\ne k$ and $\epsilon_{nkl}$ is the unit antisymmetric tensor
(the final expression in \rf{D1} can be deduced from the intermediate one
by applying twice the identity
$\LA({\bf f}\times{\bf C})\cdot\partial^2{\bf f}/\partial x_m\partial x_k\RA=0$
valid for any smooth space-periodic vector field $\bf f$ and any constant
vector $\bf C$). Hence, the following identities hold true (agreeing with
the 12 identities \rf{Di}):
$$\D^1_{22,1}=-\D^1_{33,1}=-\D^2_{21,1}=\D^3_{31,1},\quad\D^1_{23,1}=-\D^3_{21,1},\quad
\D^1_{32,1}=-\D^2_{31,1},\quad\D^2_{13,1}=-\D^3_{12,1},$$
and all the rest $\D^l_{mk,1}$ vanish.

In the next order we find
\BE\D^l_{mk,2}=\eta^{-1}(\B_m({\bf Z}_{l,2},\rZ_{k,1})
+\B_m({\bf Z}_{l,1},\rZ_{k,2}))=\eta^{-1}(\B_m({\bf Z}_{k,1},{\bf Z}_{l,2})
+\B_m({\bf Z}_{l,1},{\bf Z}_{k,2}))\EE{D2}
(the second relation \rf{rZc} has been used) and therefore
$\D^l_{mk,2}=\D^k_{ml,2}$. By \rf{re} and \rf{Za},
$${\bf Z}_{k,1}=\nabla^{-2}{\bf v}\times{\bf e}_k,\qquad{\bf Z}_{k,2}=
\nabla^{-2}\left({\bf v}\times\nabla^{-2}{\partial{\bf v}\over\partial x_k}\right);$$
for a flow $\bf v$ antisymmetric in $x_1$, ${\bf Z}_{k,1}$ is antisymmetric
in $x_1$ for $k=1$ and symmetric in $x_1$ otherwise (actually, it is simple
to show by mathematical induction that ${\bf s}_{k,j}$ feature this property
for all even $j$ and ${\bf Z}_{k,j}$ for all odd $j$); ${\bf Z}_{k,2}$ is
symmetric in $x_1$ for $k=1$ and antisymmetric in $x_1$ otherwise (moreover,
${\bf s}_{k,j}$ feature this property for all odd $j$ and ${\bf Z}_{k,j}$ for
all even $j$). Consequently, $\D^l_{mk,2}=0$ if an odd number of indices
$l,m,k$ are equal to 1. These properties of $\D^l_{mk,2}$ are in line
with identities \rf{Di} for $\D^l_{mk}$.

The asymptotics of the tensors yield the asymptotics of the quantities $Q_i$
defining the growth rates $\lambda_2^\pm$ \rf{rpm}:
\se{Qa}\be Q_1=&\,-{\eta^{-1}\over2}\,\D_{13,1}^2+{\rm O}(\eta^{-2}),\label{qa1}\\
Q_2=&\,{\eta^{-1}\over4}\Big((\D_{23,1}^1+\D_{32,1}^1)\cos2\varphi
+2\D_{33,1}^1\sin2\varphi+\D_{23,1}^1-\D_{32,1}^1\Big)+{\rm O}(\eta^{-2}),\label{qa2}\\
Q_3=&\,{\eta^{-2}\over4}\Big(2(\D_{23,2}^2+\D_{32,2}^2-\D_{13,2}^1)\cos\varphi
+2(\D_{12,2}^1-\D_{23,2}^3-\D_{32,2}^3)\sin\varphi+{\rm O}(\eta^{-1})\nonumber\\
&+\Big(2\A^1_{2,1}\D_{23,2}^3+2\A^3_{1,1}\D_{32,2}^2
-\left(\A^3_{1,1}(2\D_{32,2}^3+\D_{23,2}^3-\D_{22,2}^2)\right.\nonumber\\
&+\left.\A^1_{2,1}(2\D_{23,2}^2+\D_{32,2}^2-\D_{33,2}^3)\right)
\sin2\varphi+{\rm O}(\eta^{-1})\Big)\!\Big/\!
(\A^1_{2,1}\sin\varphi-\A^3_{1,1}\cos\varphi)\Big).\label{qa3}
\end{align}\end{subequations}

At large $\eta$ the plane consisting of the singularity points
of the eddy diffusivity tends to the limit position
\BE\lim_{\eta\to\infty}\fs=\arctan\left(
\left.\LA{\bf v}\times\nabla^{-2}{\partial{\bf v}\over\partial x_1}\RA_{\!\!3}
\ \right/\LA{\bf v}\times\nabla^{-2}{\partial{\bf v}\over\partial x_2}\RA_{\!\!1}
\ \right)+n\pi.\EE{las}
The largest in absolute value competing terms in \rf{q0} are $-\eta$ and
the singularity at $\varphi=\fs$ in $Q_3$ (unless for the two
azimuthal directions \rf{las} the numerator in $Q_3$ vanishes). For
$|\varphi-\fs|<c\eta^{-3}$, where $c$ is a sufficiently small
constant, the singularity in $Q_3$ \rf{q3} wins,
one of $\lambda_2^\pm$ becomes positive and eddy diffusivity negative.

Thus, the interaction of the oscillogenic magnetic $\alpha$-effect and eddy
diffusivity significantly enhances generation of the large-scale magnetic
fields, whose wave vectors $\bf q$ cluster near the planes $\varphi=\fs$.

\section{Numerical results for two sample flows}\label{nr}

For numerical investigation of the dynamo mechanism under consideration,
we have synthesised two sample solenoidal flows, $2\pi$-periodic in each
Cartesian variable $x_i$. One has been constructed by the following procedure:
a white-noise three-dimensional vector field is generated in the physical space
on the $128^3$-point regular grid; the antisymmetry in $x_1$ is enforced;
the field is Fourier-transformed and its mean and gradient parts are removed;
the coefficient associated with wave number $\bf k$ is divided by $2^{|\bf k|}$;
finally, the field is normalised. The energy spectrum of the resultant flow
decreases by 22 orders of magnitude. By construction, it involves all Fourier
harmonics available for the chosen resolution of $128^3$ Fourier harmonics,
and we will call it ``the full-spectrum flow''.

By contrast, the second sample flow involves only a limited (and relatively
small) number of Fourier harmonics. Six families of solenoidal flows with
a zero kinetic helicity at each point in space were introduced in \cite{RCZ}.
Their so-called family L flows are defined as
\BE{\bf v(x)}=A\nabla B-B\nabla A.\EE{fL}
When the Monge potentials $A$ and $B$ are (scalar) eigenfunctions
of the Laplacian associated with the same eigenvalue, flow \rf{fL} is
solenoidal. For $A$ odd in $x_1$ and $B$ even, flow \rf{fL} is antisymmetric
in $x_1$; the potentials are then
$$A({\bf x})=\sum_iA_i\sin\,n^{(i)}_1x_1\,
\left\{\!\!\!\begin{array}{c}\sin\\\cos\end{array}\!\!\!\right\}n^{(i)}_2x_2
\left\{\!\!\!\begin{array}{c}\sin\\\cos\end{array}\!\!\!\right\}n^{(i)}_3x_3,$$
$$B({\bf x})=\sum_iB_i\cos\,n^{(i)}_1x_1\,
\left\{\!\!\!\begin{array}{c}\sin\\\cos\end{array}\!\!\!\right\}n^{(i)}_2x_2
\left\{\!\!\!\begin{array}{c}\sin\\\cos\end{array}\!\!\!\right\}n^{(i)}_3x_3.$$
If the flow under consideration is supposed to be 2$\pi$-periodic in each
$x_i$, all wave vectors ${\bf n}^{(i)}$ in the two sums have integer components
and the same length.
We use such a flow referred to as ``non-helical'', whose potentials
are $2\pi$-periodic in each $x_i$ and are associated with the eigenvalue $-18$
of the Laplacian, where the constant coefficients $A_i$ and $B_i$ have been
generated as pseudo-random numbers uniformly distributed in the interval $[-1,1]$.
The potentials are then linear combinations of Fourier harmonics whose wave
vectors are either ($\pm3,\pm3,0$) or ($\pm4,\pm1,\pm1$), and permutations
thereof (36 wave vectors in total). Thus, for our non-helical flow the wave
vectors ${\bf n}^{(i)}$ are either (3,3,0) or (4,1,1), and their permutations;
$A$ and $B$ are comprised of 16 and 20, respectively, trigonometric monomials.

\begin{figure}[p]

\begin{subfigure}{0.5\textwidth}
\centerline{(a)\hspace*{2mm}\includegraphics[width=3in]{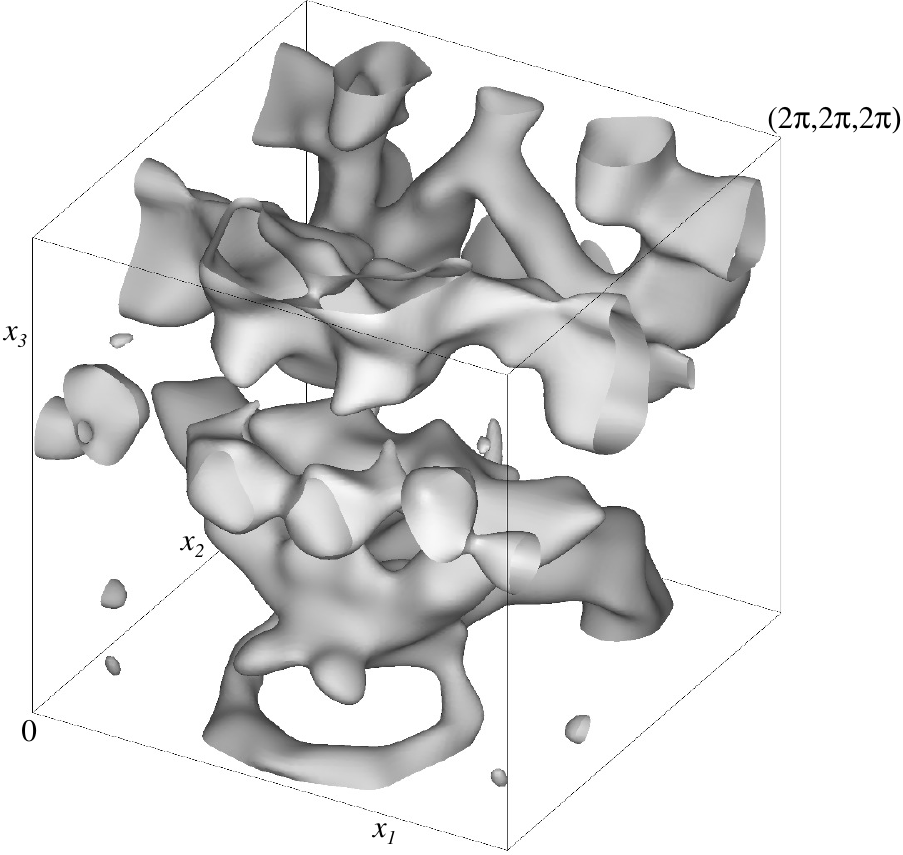}}
\end{subfigure}\begin{subfigure}{0.5\textwidth}
\centerline{(d)\hspace*{2mm}\includegraphics[width=3in]{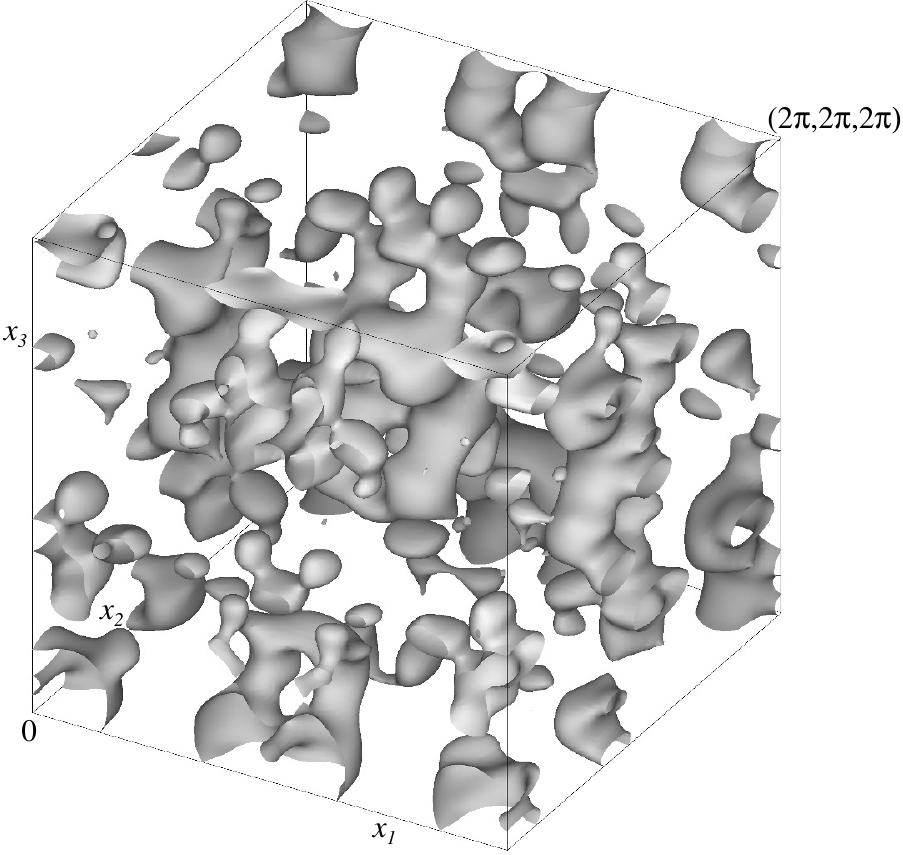}}
\end{subfigure}

\vspace*{2em}

\begin{subfigure}{0.5\textwidth}
\centerline{(b)\hspace*{2mm}\includegraphics[width=3in]{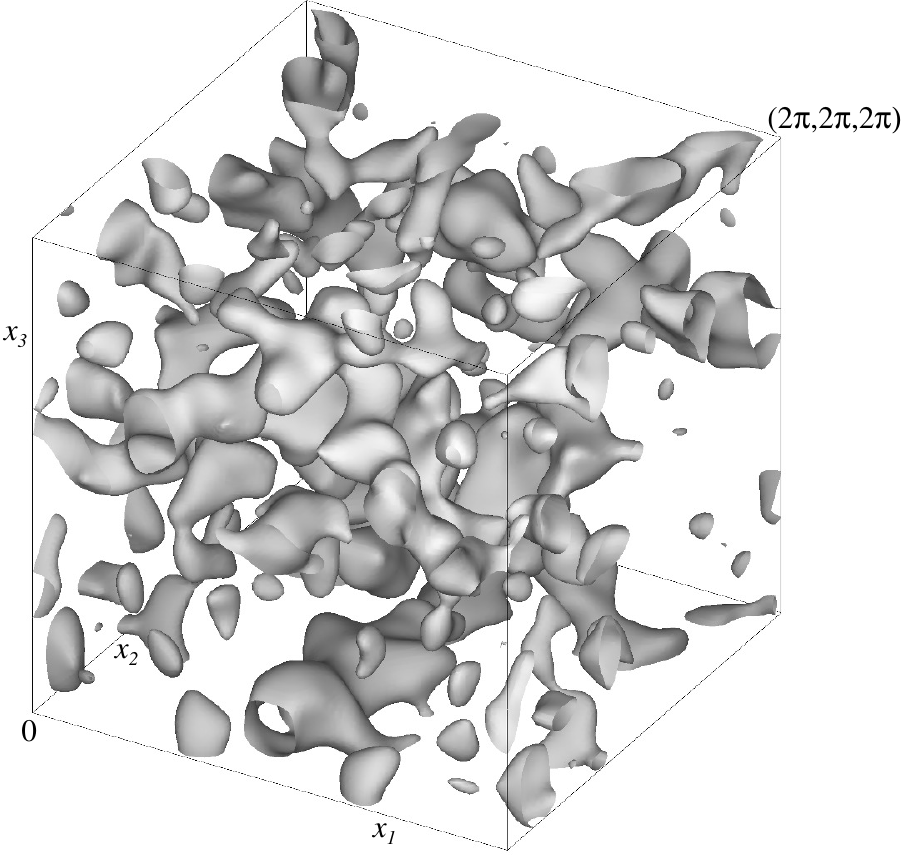}}
\end{subfigure}\begin{subfigure}{0.5\textwidth}
\centerline{(e)\hspace*{2mm}\includegraphics[width=3in]{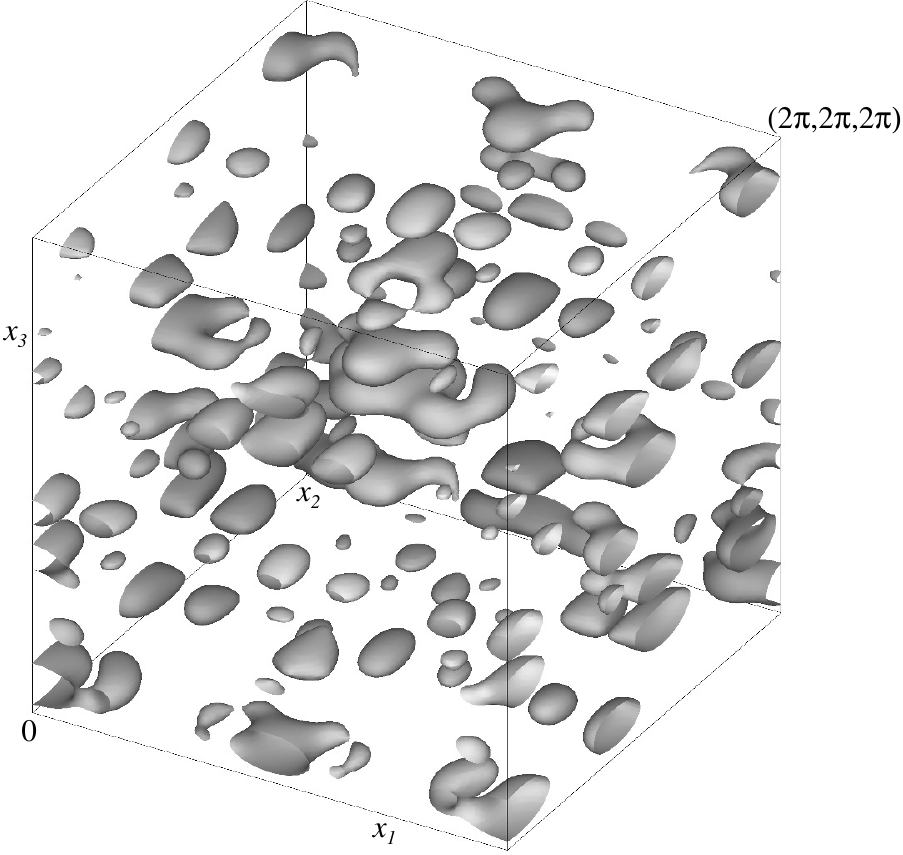}}
\end{subfigure}

\vspace*{2em}

\begin{subfigure}{0.5\textwidth}
\centerline{(c)\hspace*{2mm}\includegraphics[width=3in]{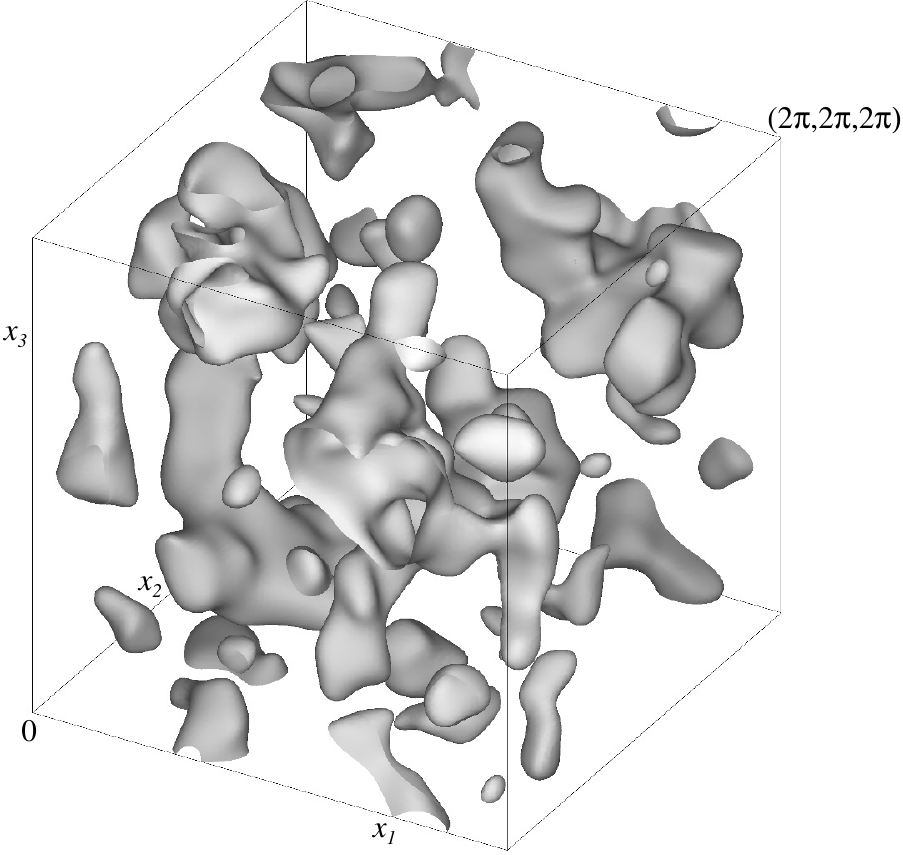}}
\end{subfigure}\hspace*{0.05\textwidth}\parbox{0.45\textwidth}{
\caption{Isosurface of the flow velocity (a), (d), vorticity (b), (e)
and kinetic helicity (c) for the full-spectrum (a)--(c) and non-helical
(d), (e) sample flows at the levels of 50\% (a), (b), 30\% (d), (e)
and 20\% (c) of the respective quantity.}\label{fv}}\end{figure}

Both fields are zero-mean and normalised so that their r.m.s.~amplitude is 1.
Let us stress that although numerical generator of pseudo-random numbers
has been used to synthesise them, both flows are steady and smooth.
The maximum flow velocities $|\bf v|$ of the full-spectrum and non-helical
flow are 2.65 and 5.30, respectively, the maximum vorticities
$|\nabla\times\bf v|$ 7.06 and 34.76,
and, for the full-spectrum flow, the kinetic helicity density
$\bf v\cdot(\nabla\times\bf v)$ maximum is 6.35\,. Isosurfaces of the velocity,
vorticity and kinetic helicity density shown in \xf{fv} attest that both flows
have an intricate structure; the non-helical flow is more spatially
intermittent than the full-spectrum one. Heuristically this may suggest
that the former flow is a better dynamo than the latter one.

\begin{table}[t]
\caption{Energy spectra decay of the computed neutral magnetic modes
${\bf s}_k$ for the two sample flows. See the text for the explanation
of the last 4 columns.}
\center\begin{tabular}{|c|c|c|c|c|r|c|r|}\hline
Flow&$\eta$&Resolution&Mode&$M$&$E_M$&$E_{\rm last}$&Decay\UP\\\hline
\parbox[t]{2mm}{\multirow{12}{*}{\rotatebox[origin=c]{90}{Full-spectrum}}}&
\multirow{3}{*}{0.02}&\multirow{3}{*}{$128^3$}
& ${\bf s}_1$ & 4 & 2.1 & $1.4\times10^{-15}$&14\UP\\
&&&${\bf s}_2$ & 3 & 3.9 & $7.6\times10^{-16}$&15\\
&&&${\bf s}_3$ & 3 & 155. & $1.5\times10^{-13}$&12\DO\\\cline{2-8}
&\multirow{6}{*}{0.01}&\multirow{3}{*}{$128^3$}
& ${\bf s}_1$ & 4 & 3.2 & $1.8\times10^{-9}$&8\UP\\
&&&${\bf s}_2$ & 3 & 11. & $1.4\times10^{-9}$&8\\
&&&${\bf s}_3$ & 3 & 471. & $3.5\times10^{-7}$&6\DO\\\cline{3-8}
&&\multirow{3}{*}{$256^3$}
& ${\bf s}_1$ & 4 & 3.2 & $2.5\times10^{-17}$&16\UP\\
&&&${\bf s}_2$ & 3 & 11. & $4.5\times10^{-16}$&15\\
&&&${\bf s}_3$ & 3 & 471. & $1.8\times10^{-17}$&16\DO\\\cline{2-8}
&\multirow{3}{*}{0.004}&\multirow{3}{*}{$256^3$}
& ${\bf s}_1$ & 5 & 9.1 & $1.1\times10^{-13}$&13\UP\\
&&&${\bf s}_2$ & 4 & 112. & $1.2\times10^{-12}$&12\\
&&&${\bf s}_3$ & 3 & 1274. & $1.0\times10^{-12}$&12\DO\\\hline
\parbox[t]{2mm}{\multirow{12}{*}{\rotatebox[origin=c]{90}{Non-helical}}} &
\multirow{6}{*}{0.02}&\multirow{3}{*}{$128^3$}
& ${\bf s}_1$ & 6 & 1.9 & $2.0\times10^{-6}$&5\UP\\
&&&${\bf s}_2$ & 3 & 7.2 & $8.7\times10^{-7}$&5\\
&&&${\bf s}_3$ & 6 & 1.1 & $1.3\times10^{-5}$&4\DO\\\cline{3-8}
&&\multirow{3}{*}{$256^3$}
& ${\bf s}_1$ & 6 & 1.9 & $2.5\times10^{-17}$&16\UP\\
&&&${\bf s}_2$ & 3 & 7.2 & $4.5\times10^{-16}$&15\\
&&&${\bf s}_3$ & 6 & 1.1 & $1.8\times10^{-17}$&16\DO\\\cline{2-8}
&\multirow{6}{*}{0.01}&\multirow{3}{*}{$128^3$}
& ${\bf s}_1$ & 6 & 24. & $2.4\times10^{-3}$&1\UP\\
&&&${\bf s}_2$ & 6 & 55. & $4.4\times10^{-3}$&1\\
&&&${\bf s}_3$ & 6 & 11. & $2.2\times10^{-2}$&1\DO\\\cline{3-8}
&&\multirow{3}{*}{$256^3$}
& ${\bf s}_1$ & 6 & 24. & $4.9\times10^{-10}$&8\UP\\
&&&${\bf s}_2$ & 6 & 55. & $1.9\times10^{-9}$&8\\
&&&${\bf s}_3$ & 6 & 11. & $6.7\times10^{-10}$&8\DO\\\hline
\end{tabular}\label{es}\end{table}

The numerically efficient procedure based on \rf{Dlk}, \rf{Dlmk}, \rf{ZS} and
\rf{xsa} has been employed for computing the eddy diffusivity tensors.
Solutions ${\bf s}_k$ to auxiliary problems of type I have been computed
by applying the code \cite{Zh} with the use of pseudo-spectral methods
(${\bf s}_k$ have the same spatial periodicity as the flow). The resolution
of $128^3$ Fourier harmonics is adequate for $\eta\ge0.02$, but becomes
insufficient for smaller $\eta$ (see table \ref{es}); for $\eta\le0.015$
the modes have been computed employing $256^3$ harmonics. As usual, in order
to verify that the computed neutral magnetic
modes are sufficiently resolved, we have computed their energy spectra, i.e.,
the quantities $E_m$ defined as the sum of squares of the moduli of the Fourier
coefficients of the mode over the wave vectors that belong to the $m$th
spherical shell ${\cal C}_m=\{{\bf n}\,|\,m-1<|{\bf n}|\le m\}$. The last four
columns in table~\ref{es} characterise how the obtained energy spectra decay:
$M$ is the number of the shell containing the maximum energy $E_M$,
$E_{\rm last}$ is the energy content in the last fully populated shell
$E_{N/2-1}$ for computations with $N^3$ harmonics, and the column ``Decay"
shows by how many orders of magnitude the spectrum decays from $10^0$,
the energy level of the inhomogeneity in the defining equation
$\L\{{\bf s}_k\}=-\L{\bf e}_k$, to the energy contained in the last fully
populated shell or to the total energy in the harmonics outside it with
the wave vectors $\bf n$ such that $|{\bf n}|\ge N/2-1$, whichever is larger.

\begin{figure}[t]
\hbox{\hspace*{-1mm}\raisebox{2in}{(a)}\hspace*{-3mm}\includegraphics[width=3.33in,height=2.6in]{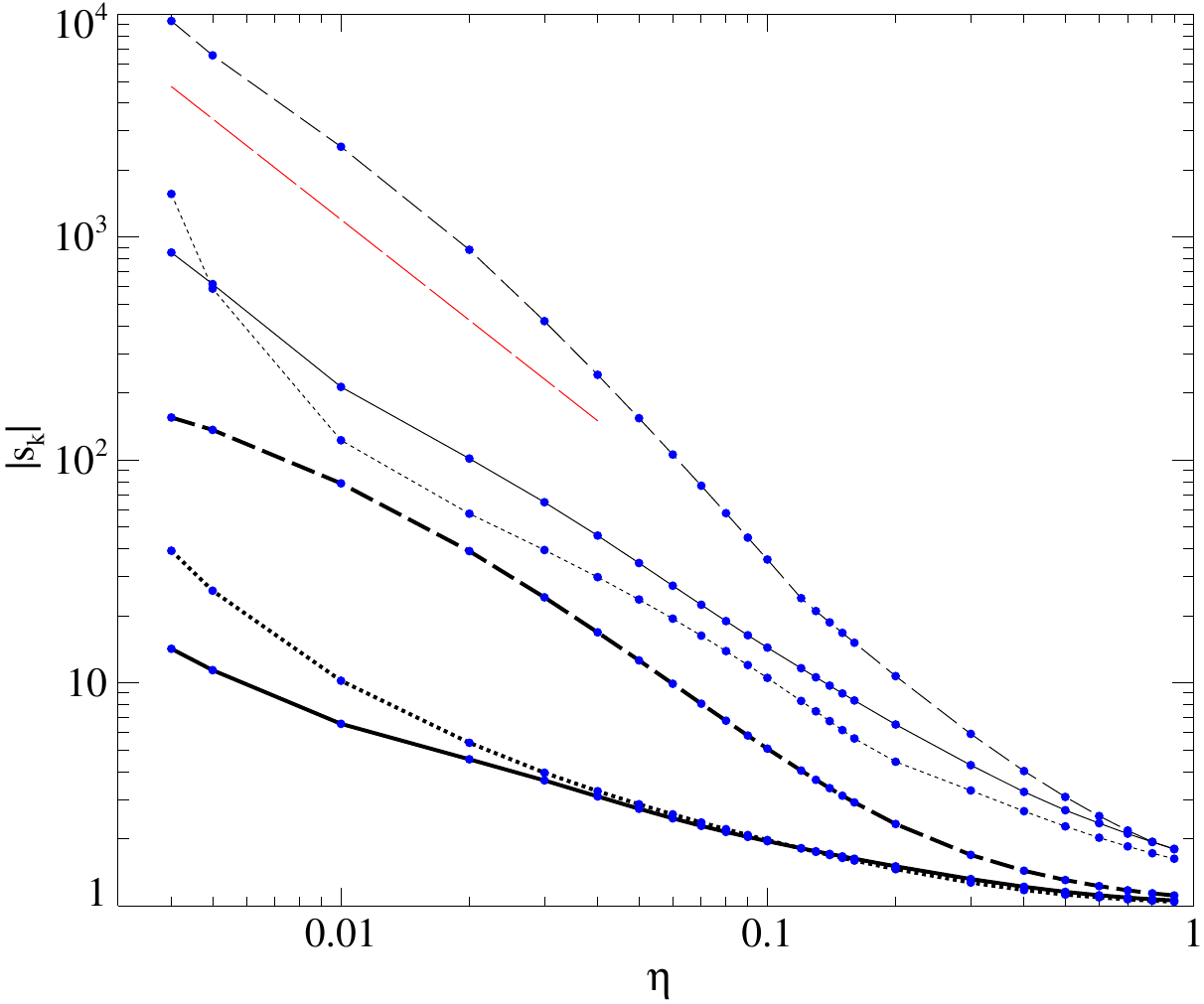}
\hspace{3mm}\raisebox{2in}{(b)}\hspace*{-3mm}\includegraphics[width=3.33in,height=2.6in]{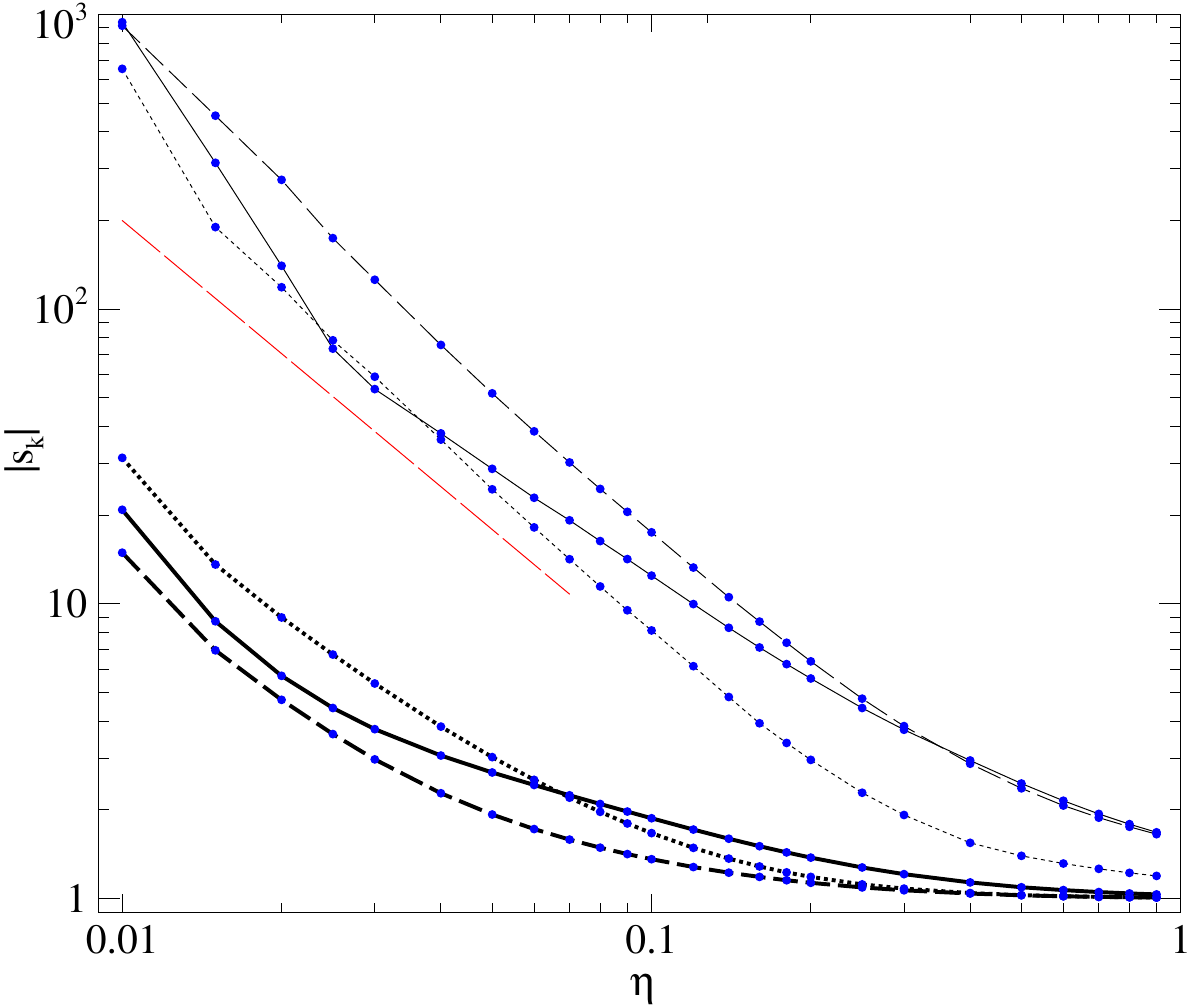}}
\caption{Maxima (thin lines) and r.m.s.~values (bold lines)
of $|{\bf s}_k|$: solid, dotted and dashed lines for $k=1,2,3$, respectively,
for the full-spectrum (a) and non-helical (b) sample flows. Solid circles
show the computed values. The long-dashed line shows, for reference,
the inclination of the plot of the function $\eta^{-3/2}$ in the log-log
coordinates of the respective figure.}
\label{fsk}\end{figure}

Figure \ref{fsk} indicates that in the limit $\eta\to0$ the maxima of
the neutral modes, ${\bf s}_k$, as well as their r.m.s.~values apparently
behave as $\eta^{-\kappa}$ for the exponent $\kappa=3/2$.
This is a tentative conjecture, since the values of $\eta$ considered here
are too high to confidently deduce the power-law behaviour from the numerical
data. An analytical derivation of this estimate for $\kappa$ is desirable.
We note that, when no small-scale dynamo operates, ${\bf s}_k$ can be obtained
by integrating the magnetic induction equation \rf{mie} with the initial
condition ${\bf h=e}_k$ up to infinite times. Such an evolutionary solution
can be described for small molecular diffusivities by M.M.~Vishik's asymptotics
\cite{Vfd}. It suggests that the maxima of ${\bf s}_k$ for $\eta\to\infty$ are
at most order $\eta^{-3/2}$. However, it is difficult to prove the asymptotics
along these lines, because that would require considering an interplay
of the asymptotics in $\eta$ with the limit of infinitely large times.

We have checked that no small-scale
dynamos operate for the magnetic molecular diffusivities employed
in our computations; thus, the large-scale dynamos considered here are not
overshadowed by (typically more efficient) small-scale dynamos.

\begin{figure}[p]
\hbox{\hspace*{-1mm}\raisebox{2in}{(a)}\hspace*{-3mm}\includegraphics[width=3.33in,height=2.6in]{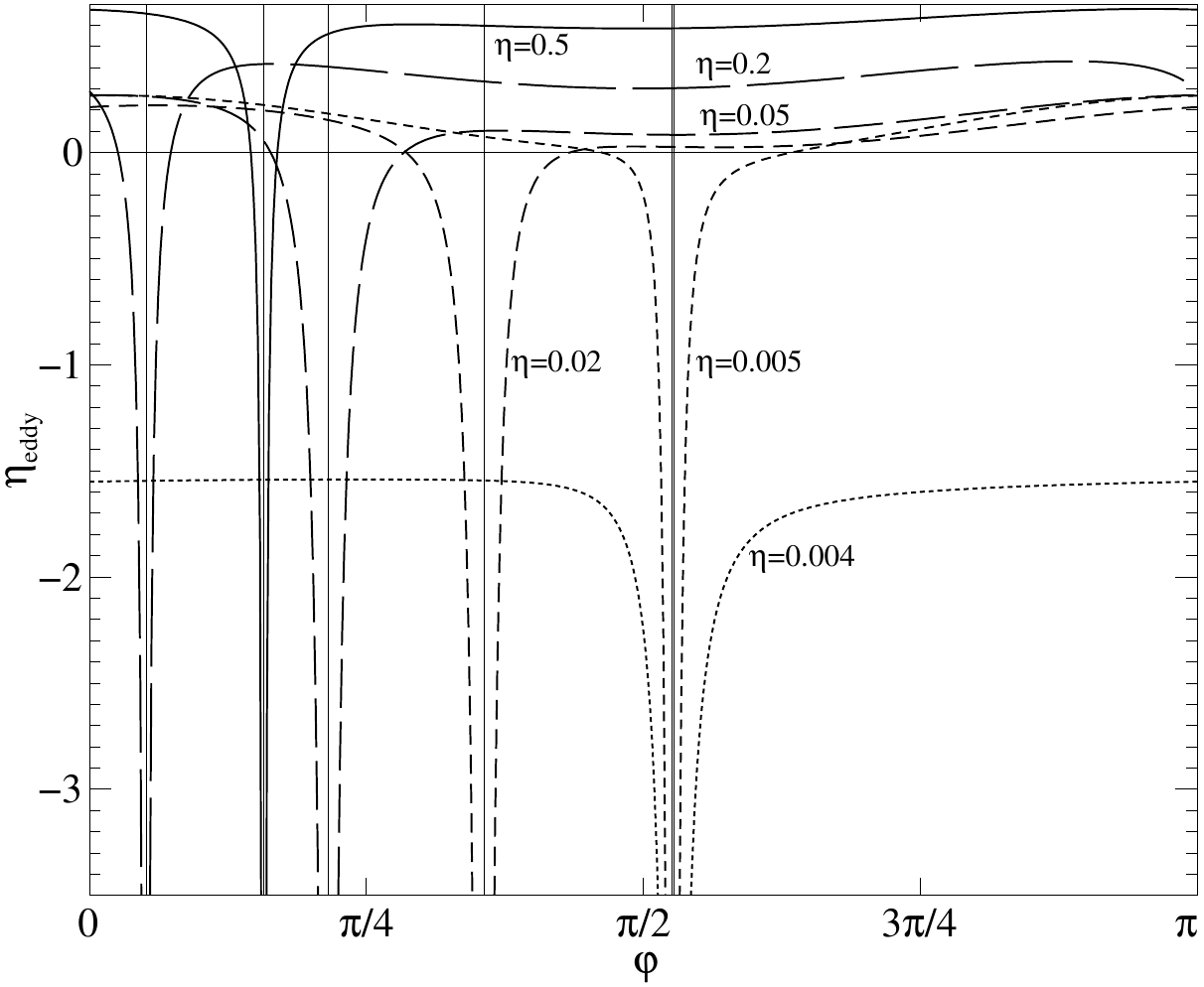}
\hspace{3mm}\raisebox{2in}{(b)}\hspace*{-3mm}\includegraphics[width=3.33in,height=2.6in]{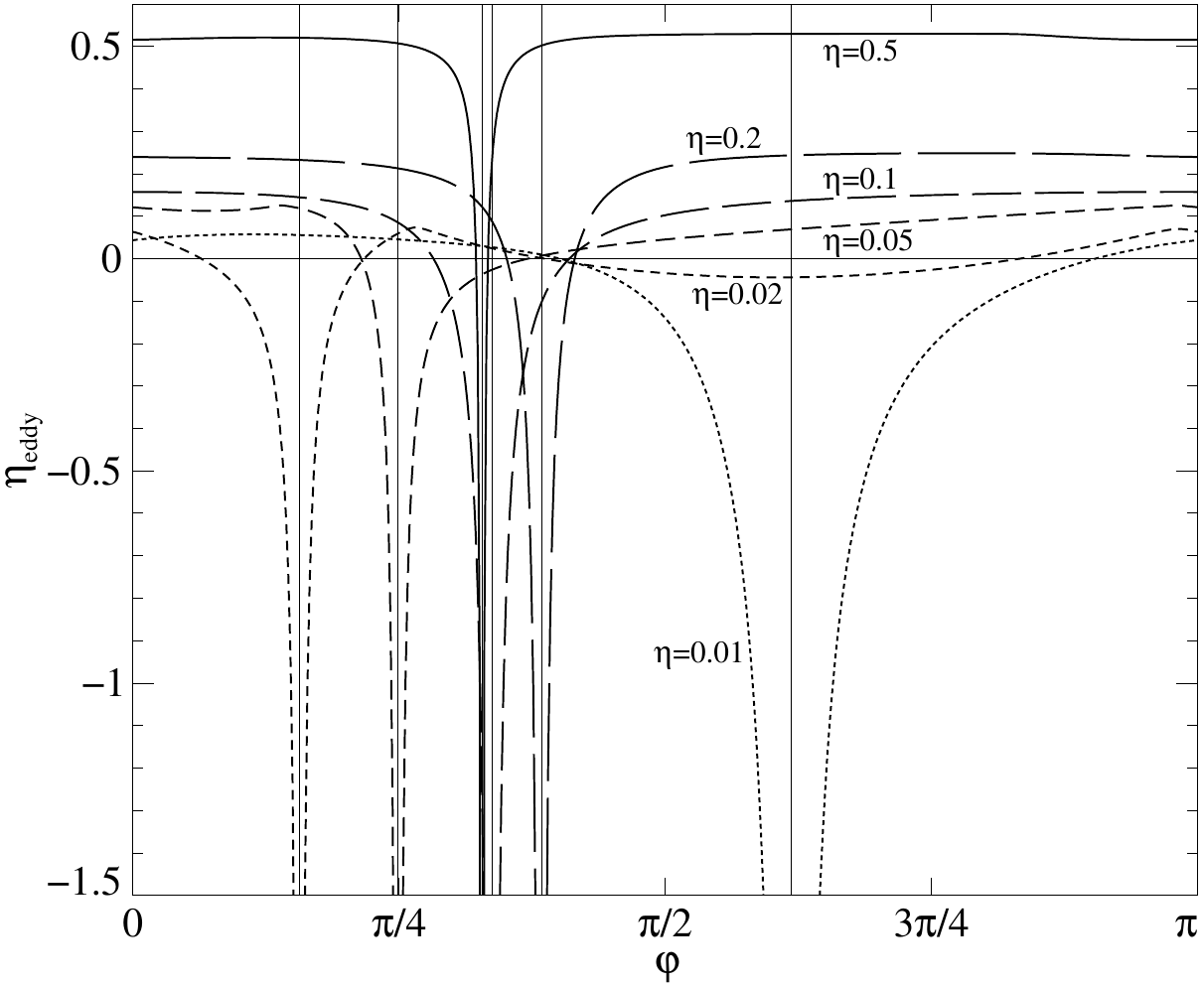}}
\caption{Minimum magnetic eddy diffusivity $\eta_{\rm eddy}(\varphi)$
for six values of magnetic molecular diffusivity (coded by the dash
length) for the full-spectrum (a) and non-helical (b) sample flows.
Thin vertical lines are the asymptotes located at the points of the singularity.}
\label{fme}\end{figure}

\begin{figure}
\hbox{\hspace*{-1mm}\raisebox{2in}{(a)}\hspace*{-3mm}\includegraphics[width=3.33in,height=2.6in]{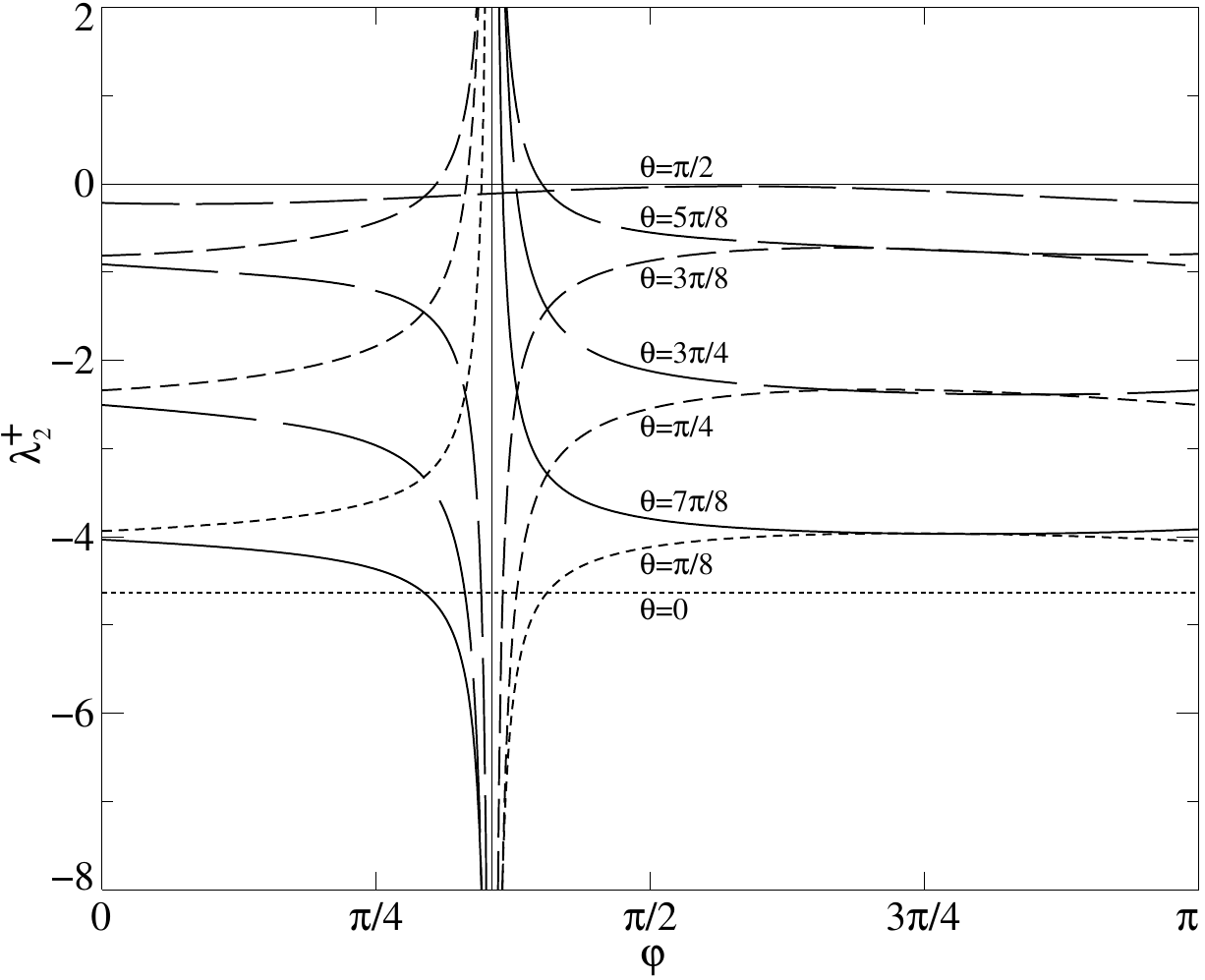}
\hspace{3mm}\raisebox{2in}{(b)}\hspace*{-3mm}\includegraphics[width=3.33in,height=2.6in]{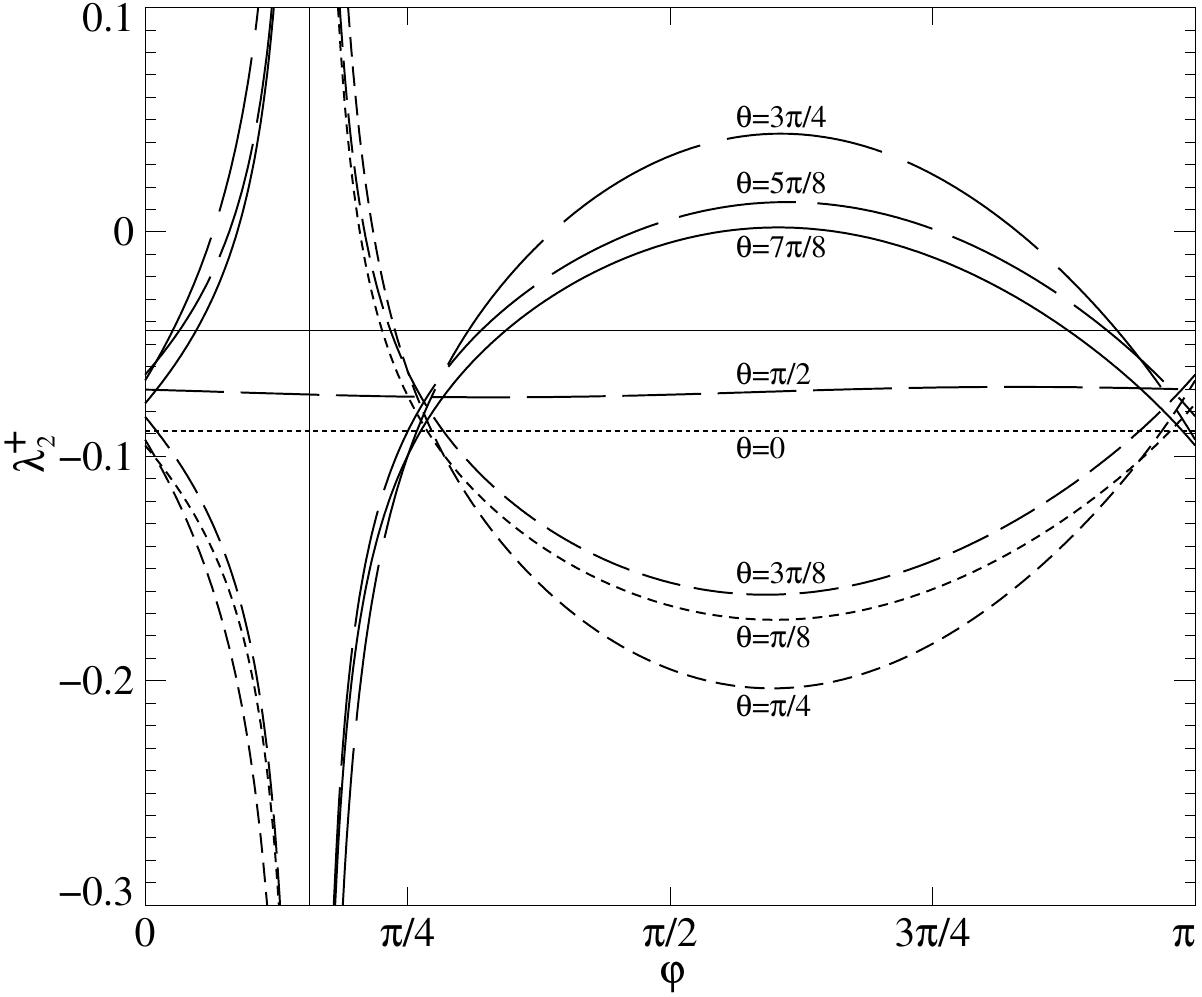}}
\caption{Growth rate $\lambda_2^+(\varphi)$ for $\eta=0.02$ for the full-spectrum
(a) and non-helical (b) sample flows. Graphs for $\theta$ (see \rf{wv}) step
$\pi/8$ are coded by the dash length.}
\label{ffi}\end{figure}

\begin{figure}
\hbox{\hspace*{-1mm}\raisebox{2in}{(a)}\hspace*{-3mm}\includegraphics[width=3.33in,height=2.6in]{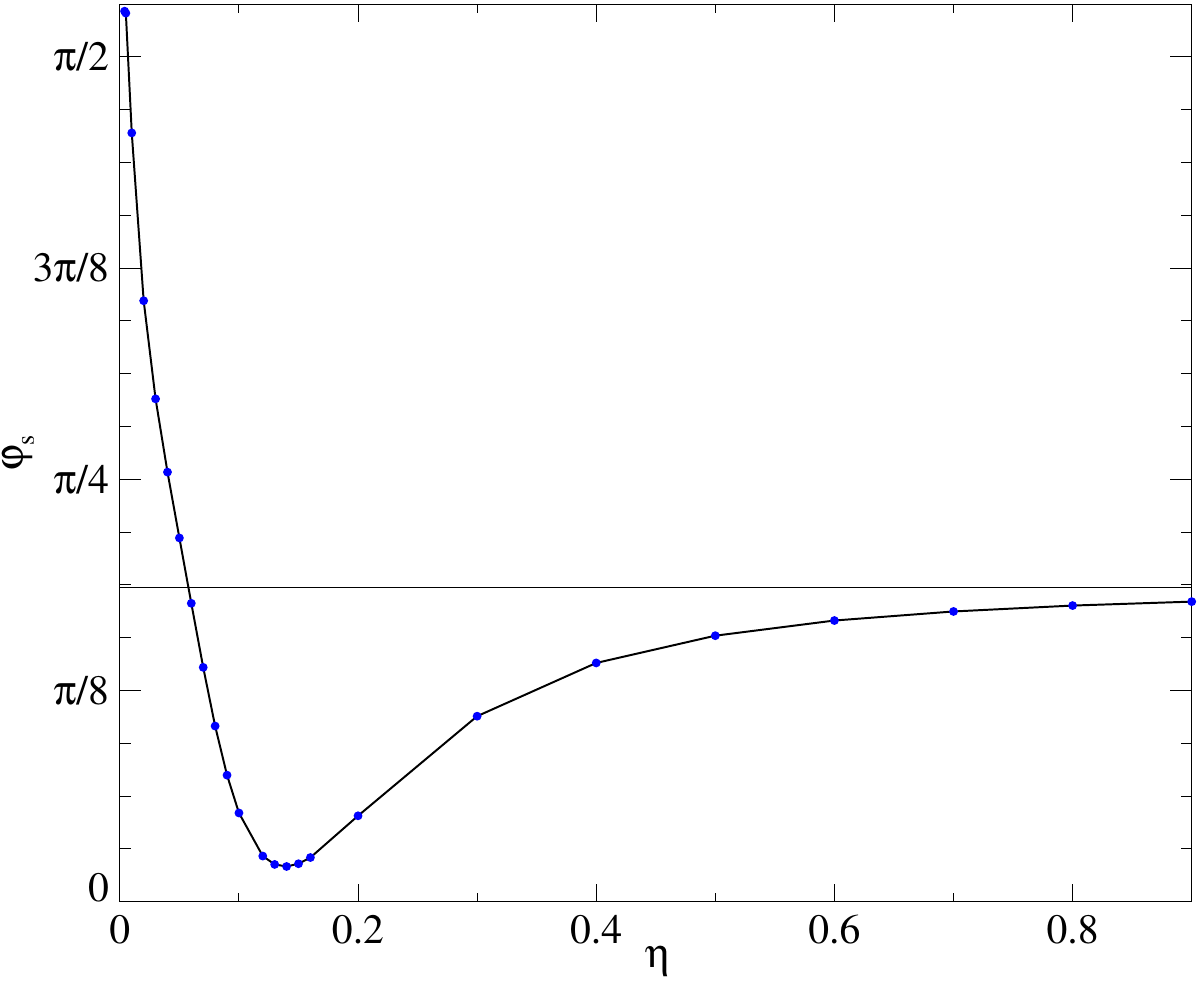}
\hspace{3mm}\raisebox{2in}{(b)}\hspace*{-3mm}\includegraphics[width=3.33in,height=2.6in]{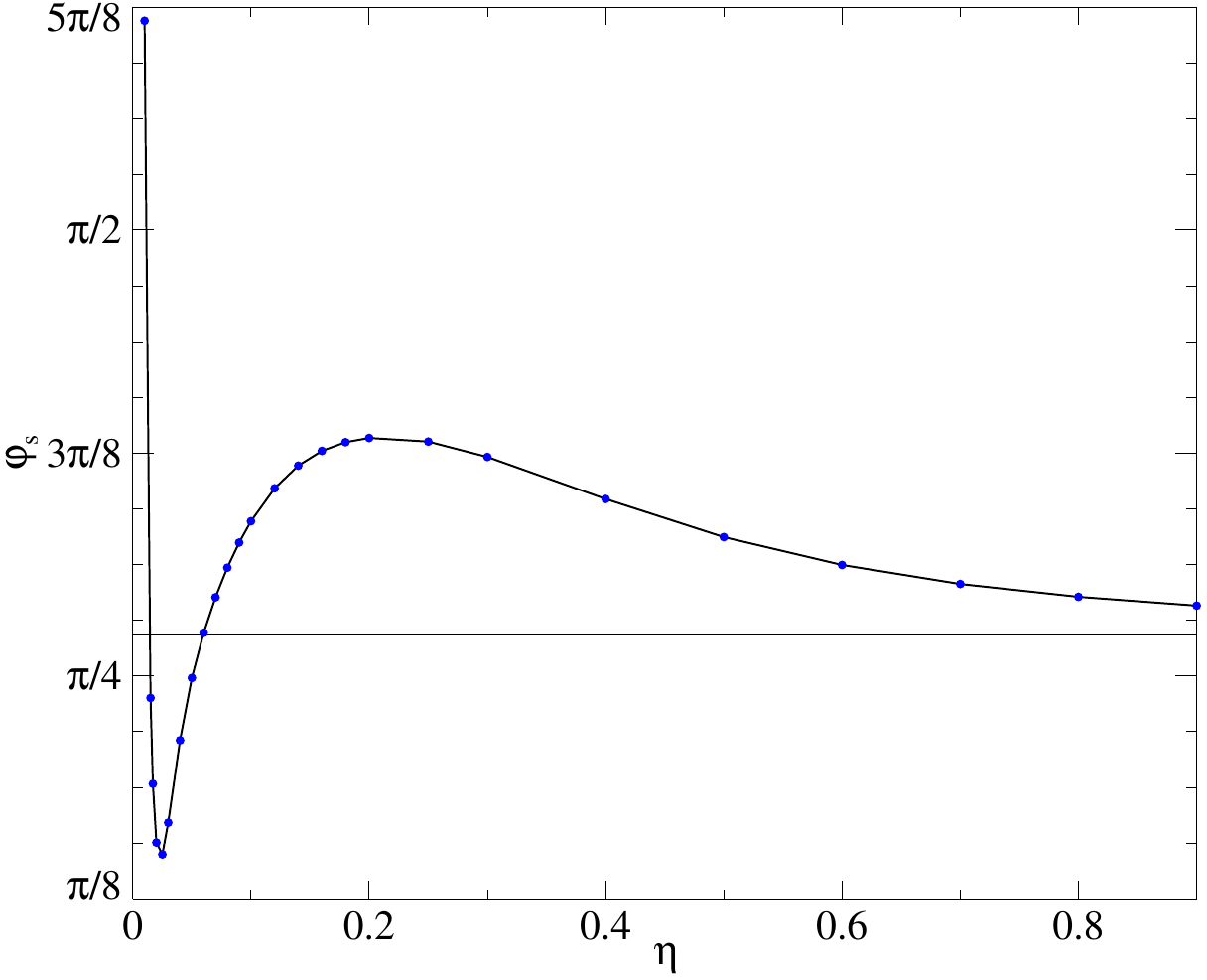}}
\caption{Location of the singularity \rf{pas}
for the full-spectrum (a) and non-helical (b) sample flows. Solid
circles show the computed values. Thin horizontal line: the limit location
of the singularity \rf{las} for $\eta\to\infty$.}
\label{fas}\end{figure}

Graphs of the computed minimum (over $\theta$ in \rf{wv}) magnetic eddy
diffusivity $\eta_{\rm eddy}(\varphi)$ as a function of the azimuthal direction
$\varphi$ of the wave vector $\bf q$ are shown in \xf{fme} for various values
of magnetic molecular diffusivity $\eta$. (Recall that the minimum eddy
diffusivity is a $\pi$-periodic function of $\varphi$, as we have demonstrated
in section \ref{gr}.) Two disjoint curves constituting a graph for a chosen
$\eta$ and separated by the vertical asymptote are shown by dashed lines
of the same dash length (depending on the $\eta$ value).

Dependencies of the growth rate $\lambda_2^+$ on $\varphi$ for latitudes
$\theta=m\pi/8$ of the wave vector $\bf q$ directions (see \rf{wv}) are shown
in \xf{ffi} for $\eta=0.02$ and integer $m$ (as usual, negative growth rates
are associated with decaying modes). Two disjoint curves related to the same
$\theta$ are shown by dashed lines of the same dash length (different
for different $\theta$). In view of the relation
$\lambda_2^+(\theta,\varphi)=\lambda_2^-(\theta,\varphi+\pi)$
similar graphs for $\lambda_2^-$ are omitted. Since $\lambda_2^+$ is invariant
under the mapping
$\varphi\mapsto\varphi+\pi$, $\theta\mapsto\pi-\theta$ (see section \ref{gr}),
we show $\lambda_2^+$ only in a half-period interval $0\le\varphi\le\pi$.
Therefore, a graph showing a certain value of $\lambda_2^+$ for $\varphi=\pi$
and some $\theta$ can be continuously extended for larger $\varphi$ by the graph
of $\lambda_2^+$ for $\theta\to\pi-\theta$ starting at $\varphi=0$ at the same
$\lambda_2^+$ value.

Both figures clearly illustrate the presence of the singularity \rf{pas}
in the term $Q_3$ in the expressions for the growth rates (see
\rf{rpm}--\rf{grm}) and minimum eddy diffusivity, as well as the positive
growth rates and
negative eddy diffusivities in the vicinity of the singularity. The higher is
$\eta$, the narrower gap between the continuous components separated
by the vertical asymptotes is observed in \xf{fme} for graphs of the minimum
eddy diffusivity, although this regularity is broken between $\eta=0.005$ and
0.02 for the full-spectrum flow in \xf{fme}(a) and between $\eta=0.05$
and 0.1 for the non-helical flow in \xf{fme}(b). Outside the gap around
the asymptotes, $\eta_{\rm eddy}$ is very insensitive to the azimuthal direction
$\varphi$.

The graphs for $\eta=0.02$ are notable: while
magnetic eddy diffusivity is negative at the interval $0.898<\varphi<1.361$
around the singularity for the full-spectrum flow, it is negative at two
longer intervals $0.202<\varphi<0.683$ around the singularity and
$1.215<\varphi<2.616$ for the non-helical flow. This indicates that
the importance of the kinetic helicity for kinematic magnetic field generation
may be overestimated (see \cite{RCZ}).

The graph of the singular direction $\fs$ in \xf{fas} shows how the direction
approaches the limit position \rf{las} when molecular diffusivity
$\eta$ indefinitely increases. Surprisingly, the high-diffusivity asymptotic
regime sets in for the molecular diffusivities as low as 1/2.

\section{Singular azimuthal directions}\label{sng}

Here we focus on wave vectors $\bf q$ belonging to the plane of singular
azimuthal directions $\varphi=\fs$ \rf{pas}. An indefinite increase
of the large-scale growth rate $\lambda_2$ suggests, as singularities often do
in physics, that for these wave vectors the considered asymptotic expansions
\rf{ps} break down. For $\varphi=\fs$, the matrix \rf{A2}
in the l.h.s.~of \rf{ei} is a $2\times2$ Jordan cell. Accordingly
\cite{Ka,Vi,Vi2}, for the singular wave vectors we seek solutions to \rf{Main}
in the form of power series in $\sqrt\varepsilon$:
\se{pss}\be{\bf h(x,X})=&\sum_{n=0}^{\infty}{\bf h}_n({\bf x,X})\varepsilon^{n/2},\label{h2}\\
\lambda=&\sum_{n=0}^{\infty}\lambda_n\varepsilon^{n/2}.\label{e2}\end{align}\end{subequations}

Substituting \rf{pss} into \rf{Main} we obtain
\BE\sum_{n=0}^\infty\left(\L{\bf h}_n
+\eta(2(\nabla\cdot\nabla_{\bf X}){\bf h}_{n-2}+\nabla_{\bf X}^2{\bf h}_{n-4})
+\nabla_{\bf X}\times({\bf v}\times{\bf h}_{n-2})
-\sum_{m=0}^n\lambda_{n-m}{\bf h}_m\right)\varepsilon^{n/2}=0;\EE{hi2}
the solenoidality of the magnetic mode again implies relations \rf{so}
for all $n\ge0$. The resultant hierarchy of equations can be solved in all
orders and provides all terms in the series \rf{pss}. They can be proved to be
asymptotic series for the solution of the large-scale dynamo problem under
consideration.

\subsection{Order $\varepsilon^0$ equation}
For $n=0$, \rf{hi2} yields \rf{eq_0}. As discussed in section \ref{O1},
the suitable solution is $\lambda_0=0$, \rf{so0}.

\subsection{Order $\varepsilon^{1/2}$ equation}
The next equation in the hierarchy \rf{hi2} is then
$\L{\bf h}_1=\lambda_1{\bf h}_0$.
Averaging yields $0=\lambda_1\LA{\bf h}_0\RA$, and for the same reasons
the relevant choice is $\lambda_1=0$. Consequently,
\BE{\bf h}_1=\sum_{k=1}^3\LA{\bf h}_1\RA_k{\bf s}_k.\EE{so1}

\subsection{Order $\varepsilon^1$ equation}
For $n=2$, \rf{hi2} implies
\BE\L{\bf h}_2+2\eta(\nabla\cdot\nabla_{\bf X}){\bf h}_0
+\nabla_{\bf X}\times({\bf v}\times{\bf h}_0)=\lambda_2{\bf h}_0.\EE{eq2}
Substituting \rf{so0} and averaging, we find
the solvability condition for this equation:
\BE\nabla_{\bf X}\times(\At\LA{\bf h}_0\RA)=\lambda_2\LA{\bf h}_0\RA,\EE{a2}
where $\At$ is the tensor of magnetic $\alpha$-effect \rf{at}.
For $\lambda_2\ne0$, \rf{so} for $n=0$ is inevitably satisfied.

We suppose that large-scale magnetic modes are Fourier harmonics \rf{fou}
for the wave vector $\bf q$ \rf{wv}, where $\varphi=\fs$. The orthogonality of
$\bf H$ and $\bf q$ justifies the use of \rf{Hpt}--\rf{qpt} for reducing
\rf{a2} to the eigenvalue problem
\BE\i\Z{\bf\Theta}=\lambda_2{\bf\Theta},\qquad{\bf\Theta}=
\left[\begin{array}{l}\Theta_{\rm t}\\\Theta_{\rm p}\end{array}\right],\EE{ei2}
where
$$\Z=\left[\begin{array}{cc}\z&\z'\\0&\z\end{array}\right],\qquad
\z=\A^2_3\cos\theta,\qquad\z'=-(\A^3_1\sin\fs+\A^1_2\cos\fs)\sin\,2\theta$$
(cf.~\rf{A2}). Generically $\Z$ is essentially a $2\times2$ Jordan cell, and
we henceforth assume $\z'\ne0$. Solutions to \rf{ei2} and \rf{eq2} are
\BE\lambda_2=\i\z,\qquad{\bf\Theta}=
\left[\begin{array}{c}1\\0\end{array}\right],\qquad
{\bf H}={\bf q}^{\rm t},\EE{H2}
\BE{\bf h}_2=\sum_{k=1}^3\left(\LA{\bf h}_2\RA_k{\bf s}_k+\e H_k\left(
\lambda_2{\bm\gamma}_k+\i\sum_{m=1}^3q_m{\bf g}_{mk}\right)\right),\EE{so2}
where ${\bm\gamma}_k({\bf x})$ and ${\bf g}_{mk}({\bf x})$ are small-scale
zero-mean space-periodic solutions to auxiliary problems of types II \rf{II}
and II$\,'$ \rf{aII}.

\subsection{Order $\varepsilon^{3/2}$ equation}
For $n=3$, \rf{hi2} yields
\BE\L{\bf h}_3+2\eta(\nabla\cdot\nabla_{\bf X}){\bf h}_1
+\nabla_{\bf X}\times({\bf v}\times{\bf h}_1)=\lambda_3{\bf h}_0+\lambda_2{\bf h}_1.\EE{eq3}
The solvability condition for this equation is obtained by
substituting \rf{so1} and averaging:
\BE\nabla_{\bf X}\times(\At\LA{\bf h}_1\RA)
=\lambda_3\LA{\bf h}_0\RA+\lambda_2\LA{\bf h}_1\RA.\EE{a3}
For $\lambda_3\ne0$, \rf{so} for $n=1$ is also inevitably satisfied.

The inhomogeneous term $\lambda_3\LA{\bf h}_0\RA$ in \rf{a3} is proportional
to $\e$ (see \rf{fou}). Together with \rf{so} for $n=1$, this implies
$$\LA{\bf h}_1\RA={\bf H}_1\e,\quad{\bf H}_1\cdot{\bf q}=0\qquad\Rightarrow\qquad
{\bf H}_1=\Theta_{\rm t1}{\bf q}^{\rm t}+\Theta_{\rm p1}{\bf q}^{\rm p},$$
whereby \rf{a3} transforms into
\BE\i\Z{\bf\Theta}_1=\lambda_3{\bf\Theta}+\lambda_2{\bf\Theta}_1,\qquad
{\bf\Theta}_1=\left[\begin{array}{l}\Theta_{\rm t1}\\\Theta_{\rm p1}
\end{array}\right].\EE{ei3}
We impose a normalisation condition $\Theta_{\rm t1}=0$ (satisfied by
multiplying the mode by a suitable linear function of $\epsilon$, without
altering the mode). Equation \rf{ei3} is then equivalent to
\BE\lambda_3=\i\z'\Theta_{\rm p1},\EE{l3}
and \rf{eq3} has a solution
$${\bf h}_3=\sum_{k=1}^3\left(\LA{\bf h}_3\RA_k{\bf s}_k+\e\left(
(\lambda_2H_{1k}+\lambda_3H_k){\bm\gamma}_k+
\i\sum_{m=1}^3q_m{\bf g}_{mk}\right)\right).$$

\subsection{Order $\varepsilon^2$ equation}
For $n=4$, we infer from \rf{hi2} the equation
$$\L{\bf h}_4+2\eta(\nabla\cdot\nabla_{\bf X}){\bf h}_2
+\eta\nabla_{\bf X}^2{\bf h}_0+\nabla_{\bf X}\times({\bf v}\times{\bf h}_2)
=\lambda_4{\bf h}_0+\lambda_3{\bf h}_1+\lambda_2{\bf h}_2.$$
Averaging, substituting \rf{so2} and recalling that $\LA{\bf h}_0\RA$ and
$\LA{\bf h}_1\RA$ are Fourier harmonics, we obtain its solvability condition:
\BE\nabla_{\bf X}\times(\At\LA{\bf h}_2\RA)+\i\e{\bf q}\times\sum_{k=1}^3
H_k\left(\lambda_2\Dw_k+\i\sum_{m=1}^3q_m\Db_{mk}\right)
=\lambda_2\LA{\bf h}_2\RA+\e(\lambda_3{\bf H}_1+(\lambda_4+\eta){\bf H})\EE{eq4}
(see \rf{Dmk}). Therefore,
$$\LA{\bf h}_2\RA={\bf H}_2\e,\qquad
{\bf H}_2=\Theta_{\rm t2}{\bf q}^{\rm t}+\Theta_{\rm p2}{\bf q}^{\rm p},$$
where we can also assume the normalisation condition $\Theta_{\rm t2}=0$.
We denote $\displaystyle{\bf\Theta}_2=\left[\begin{array}{l}\Theta_{\rm t2}\\
\Theta_{\rm p2}\end{array}\right]$.

Now \rf{eq4} takes the form
$$\i\Z{\bf\Theta}_2+\i\sum_{k=1}^3H_k\left(\lambda_2
\left[\begin{array}{r}\Dw_k\cdot{\bf q}^{\rm p}\\
-\Dw_k\cdot{\bf q}^{\rm t}\\\end{array}\right]
+\i\sum_{m=1}^3q_m\left[\begin{array}{r}\Db_{mk}\cdot{\bf q}^{\rm p}\\
-\Db_{mk}\cdot{\bf q}^{\rm t}\\\end{array}\right]\right)
=\lambda_2{\bf\Theta}_2+\lambda_3{\bf\Theta}_1+(\lambda_4+\eta){\bf\Theta}.$$
By \rf{H2} and \rf{l3}, the second component of this equation reduces to
$$-\i\sum_{k=1}^3q^{\rm t}_k\left(\z(\Dw_k\cdot{\bf q}^{\rm t})
+\sum_{m=1}^3q_m(\Db_{mk}\cdot{\bf q}^{\rm t})\right)=\z'\Theta_{p1}^2,$$
wherefrom we determine $\Theta_{p1}$ and $\lambda_3$. In view of
the symmetry properties \rf{Dti} of the tensor $\Dw$, its contribution
cancels out; using \rf{Di}, we finally find
\be\lambda_3=&\pm(1+i)\left({\z'\over2}\sin\theta\left(\D^3_{23}\cos^3\fs+
(\D^3_{33}-2\D^2_{23})\cos^2\fs\sin\fs\right.\right.\label{so4}\\
&\left.\left.+\,(\D^2_{22}-2\D^3_{32})\cos\fs\sin^2\fs+\D^2_{32}\sin^3\fs
\right)\vphantom{\z'\over2}\!\right)\!^{1/2}\!.\nonumber\end{align}

While the leading term (with the coefficient $\lambda_2$ \rf{H2})
in the expansion \rf{e2} of the eigenvalue is imaginary (i.e.,
the $\alpha$-effect remains oscillogenic for the singular azimuthal directions
$\varphi=\fs$ of the wave vector, the period of oscillations being order
$\varepsilon$, as before), the next term (with the coefficient
$\lambda_3$ \rf{so4}) has a positive non-zero real
part (unless the expression under the square root vanishes). This manifests
a large-scale dynamo operating on the time scale O($\varepsilon^{3/2}t$).
The decrease of this order for the singular directions of the wave vector
is compatible with the singular behaviour of the large-scale growth rate
$\lambda_2$ \rf{rpm} calculated for non-singular directions, for which
the large-scale dynamo operates on a slower time scale~O($\varepsilon^2t)$.

The maximum of the large-scale growth rate over $\theta$
is admitted when $|\cos\theta|=1/\sqrt3$,
\BE\max_{\theta}{\rm Re}\lambda_3=\sqrt{2\left|\D^3_{23}(\A^1_2)^3
+(\D^3_{33}-2\D^2_{23})(\A^1_2)^2\A^3_1+(\D^2_{22}-2\D^3_{32})\A^1_2(\A^3_1)^2
+\D^2_{32}(\A^3_1)^3\right|\over3\sqrt3\left((\A^3_1)^2+(\A^1_2)^2\right)}.\EE{GR}

\subsection{Numerical results}

\begin{figure}
\centerline{\hspace*{-1mm}\raisebox{2in}{(a)}\hspace*{-3mm}\includegraphics[width=3.33in,height=2.6in]{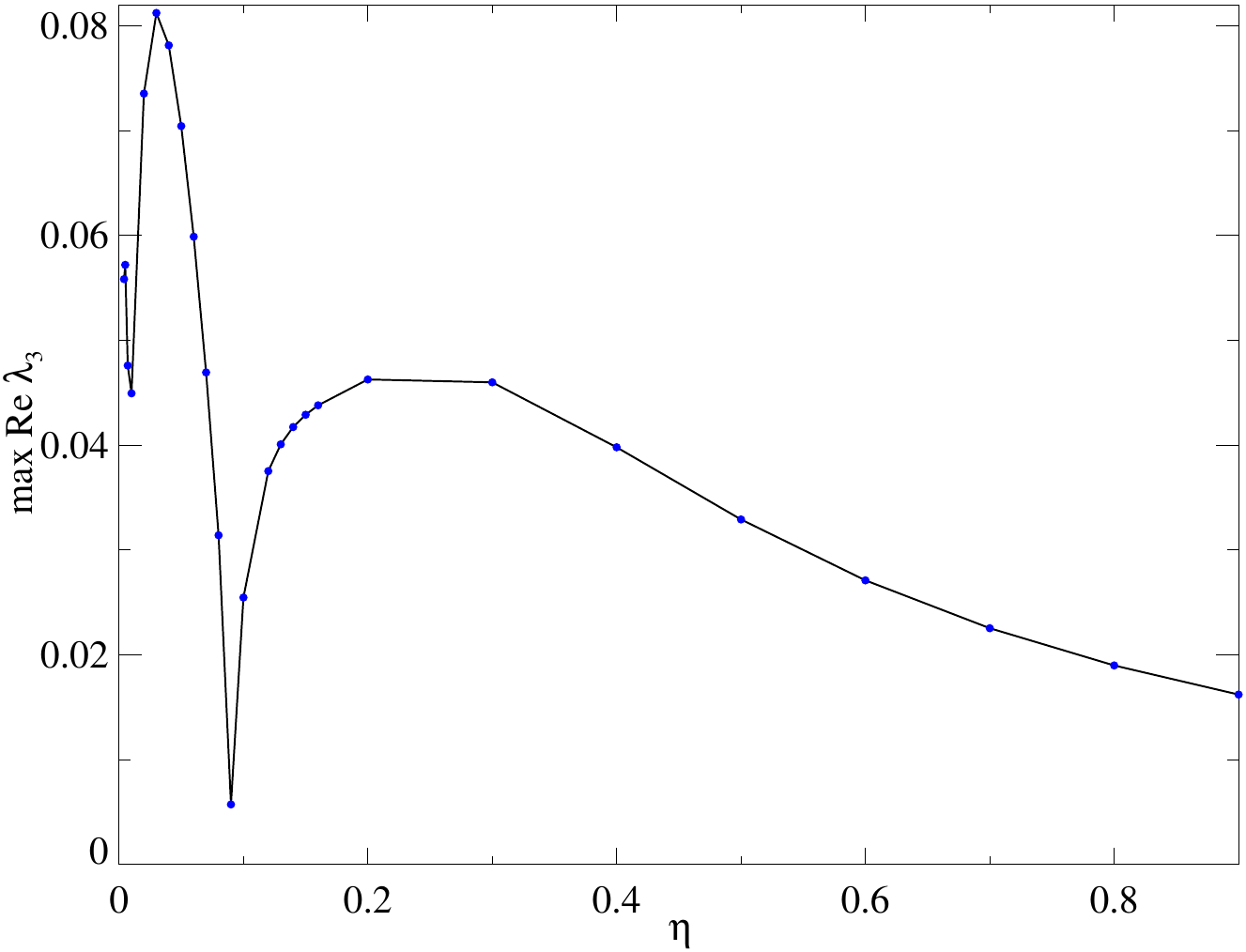}
\hspace{3mm}\raisebox{2in}{(b)}\hspace*{-3mm}\includegraphics[width=3.33in,height=2.6in]{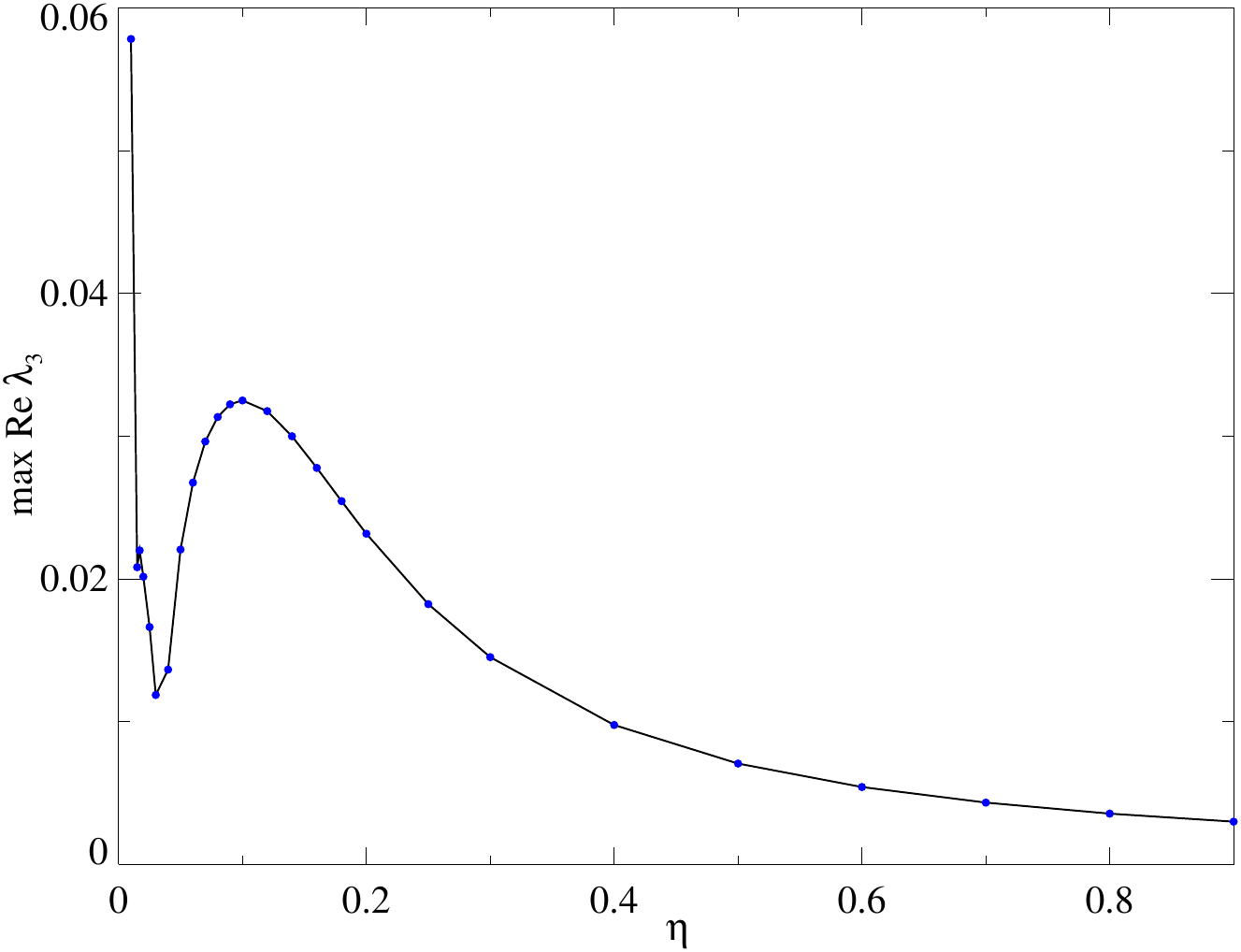}}
\caption{Maxima \rf{GR} of growth rates on the time scale O($\varepsilon^{3/2}t$)
over wave vectors $\bf q$ belonging to the plane of singular azimuthal
directions $\varphi=\fs$ for the full-spectrum (a)
and non-helical (b) sample flows. Solid circles show the computed values.}
\label{x32}\end{figure}

Maxima \rf{GR} of growth rates on the time scale O($\varepsilon^{3/2}t$)
over wave vectors $\bf q$ constituting the plane of singular azimuthal
directions $\varphi=\fs$ have been computed for varying molecular diffusivity
(see \xf{x32}) for the same two sample flows as in section \ref{nr}. We have
done computations with the resolution of $128^3$ Fourier harmonics. For
the smallest values of $\eta$ (0.004, 0.005 and 0.01 for the full-spectrum
flow, \xf{x32}(a), and 0.01 for the non-helical flow, \xf{x32}(b))
the maxima have been verified in computations with the resolution of $256^3$
harmonics; they proved to coincide with the results of the $128^3$-harmonics
runs in 4 significant digits.

The behaviour of the maximum growth rates for small molecular diffusivities
becomes intermittent. Near the points, where the maximum vanishes, it exhibits
a non-smooth (albeit continuous) behaviour, having in, agreement with \rf{GR},
square-root-like cusps (although it is not sufficiently well resolved at the scale
of \xf{x32}). Such points are detected numerically as points, where there is
a change of the sign of the cubic expression, whose absolute value is taken
in the numerator in \rf{GR}, and they have been evaluated by interpolation.
For the non-helical flow, the growth of $\max_{\theta}{\rm Re}\lambda_3$
for decreasing $\eta$ is a consequence of the onset of the small-scale
dynamo action; the maximum tends to infinity, when the critical molecular
diffusivity for the onset of generation is approached. The mathematical reasons
for such a behaviour are the same as the infinite decrease of magnetic eddy
diffusivity (obtained by developing the conventional expansions \rf{ps}),
when the dominant eigenvalue of the small-scale magnetic induction operator
vanishes (see \cite{VZ}) and solutions to auxiliary problems become large
in magnitude.

We have observed in section \ref{nr} that the non-helical flow is more
spatially intermittent than the full-spectrum one, suggesting that
the former flow is a better dynamo than the latter one. However, examination
of \xf{x32} reveals that the maximum growth rates \rf{GR} for
the full-spectrum flow are higher than those for the non-helical one,
except for when the critical molecular diffusivity for the onset of generation
by the non-helical flow is approached (and outside a small interval
in the vicinity of the zero maximum growth rate for the full-spectrum flow).

\section{Concluding remarks}\label{con}

We have presented a two-scale dynamo sustained by the simultaneous
action of the two most important large-scale mechanisms: the magnetic
$\alpha$-effect and negative eddy diffusivity. (Interaction of the two
mechanisms has somewhat amplified the entanglement of algebra
involved in application of the homogenisation techniques. For instance,
while auxiliary problems of type II \rf{II} coincide with those encountered
in the standard case of emergence of the phenomenon of magnetic eddy
diffusivity when the magnetic $\alpha$-effect is absent (see Chapter 3
in \cite{VZ}), auxiliary problems of type II$\hspace{1pt}'$ \rf{aII} have
no analogues.) The influence of the $\alpha$-effect
on generation of large-scale field is intricate. By virtue of \rf{l2},
the growth rates $\lambda_2^\pm$ depend on the entries of the $\alpha$-effect
tensor in two ways: via the first term, $\lambda_1^\pm$, in the expansion
of the eigenvalue $\lambda$, and via the dependence of the ratios of components
of vectors \rf{Hpm} and \rf{Tpm} on the ratio $\A^3_1/\A^1_2$.
Not surprisingly, both the $\alpha$-effect and eddy diffusivity have nothing in
common with the total kinetic helicity $\int{\bf v}\cdot(\nabla\times\bf v)\d\bf x$;
this becomes especially transparent when considering the large $\eta$ limit
(see \rf{ata}--\rf{D1}); a heuristic argument explaining this in terms of
the flow complexity and various topological properties of knottedness
of vorticity lines was put forward in \cite{RCZ}.

The dynamo operates as follows. The $\alpha$-effect creates a large-scale order
$\varepsilon^0$ mean field $\LA{\bf h}_0\RA$ (see \rf{av1}),
oscillating in time on the time scale O$(\varepsilon^{-1})$,
that neither grows, nor decays on this time scale. This mean field is
accompanied by an O(1) suite field $\{{\bf h}_0\}$ fluctuating in space (see
\rf{so0}), from which the small-scale flow creates an O$(\varepsilon)$
fluctuating field $\{{\bf h}_1\}$ (see \rf{sol_1}--\rf{aII}). Its interaction
with the flow gives rise to an O$(\varepsilon)$
mean e.m.f.~$\LA{\bf v}\times\{{\bf h}_1\}\RA$ resulting in emergence
of the magnetic eddy diffusivity, that can sustain the growth of the mean
field $\LA{\bf h}_0\RA$ on the O$(\varepsilon^{-2})$ time scale. Thus both
the $\alpha$-effect and magnetic eddy diffusivity emerge due to the interaction
of the small-scale flow with various components of the large-scale fluctuating
magnetic field. The physics behind the two effects being basically the same,
the difference between them is clearly observed at the mathematical level:
the $\alpha$-effect acts on the O$(\varepsilon^{-1})$ time scale and it is
described by the $\alpha$-effect operator
$${\bf h}\mapsto\nabla_{\bf X}\times(\At\bf h)$$
(cf.~\rf{aleq}) which is a differential operator of the first order;
eddy diffusivity acts
on the O$(\varepsilon^{-2})$ time scale, and the eddy diffusivity operator
$${\bf h}\mapsto\eta\nabla^2_{\bf X}{\bf h}+\nabla_{\bf X}\times\sum_{k=1}^3
\left(\lambda_1h_k\Dw_k+\sum_{m=1}^3\Db_{mk}{\partial h_k\over\partial X_m}\right)$$
(cf.~its symbol with the l.h.s.~of \rf{h_eq}) is a differential operator
of the second order. We encounter here a new $\alpha$-effect-like term
$\lambda_1\nabla_{\bf X}\times\sum_{k=1}^3h_k\Dw_k$, which
does not appear in the eddy diffusivity operator for parity-invariant flows.
(Since $\lambda_1$ \rf{lpm} is linear in $\bf q$, the contribution of this term
is also quadratic in $\bf q$, and hence it can also be regarded as
a second-order operator.)

Astrophysical dynamos are running at high kinetic, Re, and magnetic, Rm,
Reynolds numbers such that the magnetic Prandtl numbers ${\rm P}_m={\rm Rm/Re}$
are typically very small (${\rm P}_m\ll1$, e.g., in planetary interiors)
or large (${\rm P}_m\gg1$, e.g., in the interstellar medium). Such dynamos
are problematic for theoretical analysis (see \cite{Sch,Br,Mo}
for a discussion). An attractive feature of the two-scale dynamo
under consideration, perhaps making it useful for astrophysical applications,
is that (generically) it generates a mean field for all magnetic molecular
diffusivities, i.e., for all ${\rm P}_m$ (although it is unclear what might
coerce the small-scale turbulence to sustain the antisymmetry necessary
for this dynamo). Apparently, the large-scale dynamo mechanism
under consideration is not significantly hindered by the $\alpha$-quenching,
since both the numerator and denominator in the singular function
$Q_3(\varphi)$ \rf{q3} --- a constituent part of the growth
rate $\lambda_2$ \rf{rpm} --- are linear and homogeneous in the entries
of the $\alpha$-effect tensor. In the maximum growth rate \rf{GR} for
singular azimuthal directions, the numerator and denominator are also
both homogeneous in the entries of the $\alpha$-effect tensor, but their orders
are different; as a result, $\max_{\theta}{\rm Re}\lambda_3\to0$
when the $\alpha$-quenching occurs.

Our dynamo is characterised by a strong spatial focusing: interaction
of the oscillogenic $\alpha$-effect and eddy diffusivity results in an upsurge
of large-scale magnetic fields whose wave vectors belong to the ``singular''
plane $\varphi=\fs$ (normal to the plane of mirror antisymmetry of the flow).
The respective mean fields are predominantly toroidal (i.e., predominantly
parallel to the plane of mirror antisymmetry); they oscillate
in the slow time O($\varepsilon t$) and have finite growth rates in the slow
time O($\varepsilon^{3/2}t$). They are thus generated faster than fields
for wave vectors outside the singular plane, whose growth rates are finite
in the slow time O($\varepsilon^2t$).

The following questions remain open and will be addressed in future work.\\
$i$. The asymptotics that we have considered is in $\varepsilon\to0$ followed
by $\eta\to\infty$. It is of interest to consider other branches
of solutions, in which $\varepsilon$ and $\eta$ vary simultaneously.\\
$ii$. Our small-scale flow is supposed to mimick the small-scale turbulent
motion, making it is desirable to consider a more realistic case of flow,
periodic in fast time (following the developments in chapter 4 of \cite{VZ}).\\
$iii$. For a flow, antisymmetric in one Cartesian variable, that we have
chosen to study, the $\alpha$-effect is oscillogenic for all magnetic molecular
diffusivities, and also the eddy diffusivity tensors \rf{Dmk} possess various
symmetry-type properties (\rf{Dti} and \rf{Di}) which has given an opportunity
to considerably simplify equations (e.g., expression for the growth rate
$\lambda_2$ \rf{rpm}). It was demonstrated in \cite{RCZ} that a flow features
an oscillogenic $\alpha$-effect, whenever the symmetrised $\alpha$-effect
tensor (whose entries are $(\A^m_k+\A^k_m)/2$) has one zero eigenvalue
and the two remaining eigenvalues have opposite signs (in our case the two
non-zero eigenvalues of the symmetrised $\alpha$-effect tensor are
$\pm\sqrt{(\A^1_3)^2+(\A^2_1)^2}$). An interesting question is to characterise
the class of flows, for which the intermediate eigenvalue of the symmetrised
$\alpha$-effect tensor is zero for all molecular diffusivities, and to derive
for such flows the main term in the expansion of the growth rate Re\,$\lambda_2$.

\section*{Acknowledgements}

RC was supported by the project POCI-01-0145-FEDER-006933/SYSTEC
(Research Center for Systems and Technologies, University of Porto) financed
by FEDER (Fundo Europeu de Desenvolvimento Regional / European Regional
Development Fund) through COMPETE 2020
(Programa Operacional Competitividade e Internacionaliza\c c\~ao), and by FCT
(Funda\c c\~ao para a Ci\^encia e a Tecnologia, Portugal). VZ was partially
supported by CMUP (Centro de Matem\'atica da Universidade do Porto,
UID/MAT/00144/2019), which is funded by FCT with national
(MCTES) and European structural funds through the programs FEDER under
the partnership agreement PT2020, and projects STRIDE
[NORTE-01-0145-FEDER-000033] funded by FEDER -- NORTE 2020 and MAGIC
[POCI-01-0145-FEDER-032485] funded by FEDER via COMPETE 2020 -- POCI.
The authors would like to thank the anonymous Referee, whose comments
have helped us to improve the paper.

\end{document}